\documentclass[12pt,amsmath,jcp,superscriptaddress,preprint]{revtex4-1}

\usepackage{color}
\usepackage{amsfonts}
\usepackage{amssymb}
\usepackage{graphicx}

\begin{document}
\title{Electronic friction in interacting systems}
\author{Feng Chen}
\affiliation{Department of Physics, University of California San Diego, La Jolla, CA 92093, USA}
\author{Kuniyuki Miwa}
\affiliation{Department of Chemistry \& Biochemistry, University of California San Diego, La Jolla, CA 92093, USA}
\author{Michael Galperin}
\email{migalperin@ucsd.edu}
\affiliation{Department of Chemistry \& Biochemistry, University of California San Diego, La Jolla, CA 92093, USA}

\begin{abstract}
We consider effects of strong light-matter interaction on electronic friction in molecular junctions
within generic model of single molecule nano cavity junction. 
Results of the Hubbard NEGF simulations are
compared with mean-field NEGF and generalized Head-Gordon and Tully approaches.
Mean-field NEGF is shown to fail qualitatively at strong intra-system interactions, while 
accuracy of the generalized Head-Gordon and Tully results is restricted to situations of well separated intra-molecular excitations,
when bath induced coherences are negligible.
Numerical results show effects of bias and cavity mode pumping on electronic friction.  
We demonstrate non-monotonic behavior of the friction on the bias and intensity of the pumping field
and indicate possibility of engineering friction control in single molecule junctions.  
\end{abstract}

\maketitle

\section{Introduction}\label{intro}
Dynamics of open quantum systems is an active area of reasearch due to its
fundamental complexity and promise of technological applications.
In single molecules and single molecule junctions, many studies forcus 
on dynamics caused by interactions between electronic and vibrational 
degrees of freedom in the molecule. In particular, the interactions are central
to spectroscopy~\cite{MukamelJCTC16,SubotnikJCTC15,SubotnikJCP14,PetitSubotnikJCP14}, 
bias-induced~\cite{GuoJPCL16,SubotnikJPCC15} 
and photo-chemistry~\cite{LeiZhuJPhotoChemPhotoBiolA16}, 
electron ~\cite{PrezhdoBeljonnePCCP15} and
energy~\cite{Imada2017,Imada2016,CokerAnnRevPhysChem16,TullyWodtkeShaferJPCC15} transfer,
coherent control~\cite{HenricksenJCP16}, 
radiative~\cite{ourNL}
and non-radiative~\cite{LongJPCL16}  electronic relaxation, 
and stability of junctions~\cite{nitzan_kinetic_2018,ThossPeskinNL18,FortiVazquezJPCL18,SegalPCCP12,BrandbygePRL11}.
Understanding mechanisms of the interactions and developing 
theoretical description is crucial for engineering 
optoelectronic~\cite{OhmuraAIPAdv16,TretiakAccChemRes14}
and optomechanical~\cite{FilatovJPCL16} molecular devices.

Current-induced nuclear forces is an important part of such considerations
directly related to study of dynamics in open nonequilibrium
molecular systems under assumption of time scale separation between
electronic and nuclear degrees of freedom (the Born-Oppenheimer approximation).
The time scale separation allows formulation of the stochastic Langevin
equation for classical molecular nuclei driven by quantum 
nonequilibrium electronic subsystem. Electronic degrees of freedom 
induce renormalization of the adiabatic nuclear potential
and lead to appearance of electronic friction and stochastic forces. 
The latter two are related by the fluctuation-dissipation theorem.
Roughly, theoretical derivations can be separated to exact considerations
within path-integral~\cite{NewnsPRB95,TodorovPRB12,HedegardBrandbygePRL15} or scattering~\cite{VonOppenPRB12,vonOppenPRL11,vonOppenBJN12} approaches 
and to more qualitative formulations employing quantum-classical  Liouville equation~\cite{ShenviTullyFaradDisc12,SubotnikJPCB14,NitzanSubotnikJCP15,SubotnikJCP16,SubotnikPRL17,SubotnikPRB17}. 
Some formulations of electronic friction even rely on Golden rule type
derivations~\cite{TullyPRL16,TullyPRB16}.
In terms of accounting for interactions within the electronic subsystem,
exact considerations were mostly restricted to mean-field level of treatment.
Recently, more general exact considerations started to appear
capable of taking into account intra-molecular interactions within
perturbative diagrammatic expansion in the interactions strengths~\cite{hopjan_molecular_2018,kantorovich_nonadiabatic_2018}. 
However, strong intra-system interactions are beyond capabilities
of these approaches.

Recently, we formulated general derivation of the current-induced
nuclear forces applicable in nonequilibrium molecular systems
with arbitrary intra-system interactions~\cite{chen_current-induced_2019}. 
The derivation is valid
for any strength and form of interactions in electronic subsystem 
and/or between electronic and nuclear degrees of freedom.  
We showed that electronic friction can be expressed in terms of
retarded projections of single-particle nonequilibirum Hubbbard 
Green function. The Hubbard NEGF - recently introduced
by us many-body flavor of nonequilibrium Green function method~\cite{ChenOchoaMGJCP17} -
appears to be reasonably accurate in a wide range of parameters~\cite{MiwaChenMGSciRep17}. 
We note in passing that formulations of electronic
friction in standard NEGF results in two-particle Green function -
an object much harder to handle than single-particle result of
the Hubbard NEGF. 
The Hubbard NEGF  uses many-body molecular states as a basis. 
Thus, all intra-system interactions are taken into account exactly. 
In this respect it is similar to the quantum master equation (QME) formulations.
However, contrary to standard QME (such as, e.g., Lindblad/Redfield QME), 
the Hubbard NEGF is a diagrammatic expansion in the system-bath(s) 
coupling(s), which means that under a particular order in expansion 
one sums all diagrams of this order. Thus, the methodology overcomes
usual restrictions ($k_BT\gg\Gamma$, where $k_BT$ is thermal energy and
$\Gamma$ electron escape rate) of the QME schemes and is capable 
of accounting for non-Markov character of system-bath dynamics.
  
Here, we apply the Hubbard NEGF to study effects of strong intra-system 
interactions on electronic friction. In particular, we consider 
single-molecule cavity junction and discuss effects of strong
light-matter (plamson-molecular exciton) interaction on
electronic friction. We note that strong light-matter interaction in 
single molecule junctions was recently demonstrated experimentally 
in scanning tunneling microscope-induced plasmonic nanocavities~\cite{kongsuwan_suppressed_2018,chikkaraddy_single-molecule_2016}. 
While so far only optical response of the junction has been studied, similar measurements in
current-carrying molecular junctions will probably become a reality in the nearest future.

Structure of the paper is the following. In Section~\ref{model} we introduce 
a model of molecular junction and give brief introduction to the Hubbard NEGF
and to way of simulating electronic friction. 
Numerical results and discussion are presented in Section~\ref{numres}.
Section~\ref{conclude} summarizes our findings and outlines goals for 
future research.

\section{Model and method}\label{model}
We consider junction which consists of a molecule in nanocavity
coupled to metallic contacts and to external radiation field.
Molecule is modeled as two-level system, $\varepsilon_m$ ($m=1,2$) 
with electron hopping $t$ between the levels
and with coupling to two vibrational modes, $Q_1$ and $Q_2$.
The modes will be treated classically - we will be interested
in electronic friction (dissipative part of electronic force)
acting on the nuclear motion.
Nanocavity is represented by single cavity mode modeled as 
harmonic oscillator of frequency $\omega_{c}$. 
The mode is coupled to molecular exciton,
modeled as transition between the two levels, and to external radiation field.
Two contacts, $L$ and $R$, are modeled as free electron reservoirs,
each at its own equilibrium. Radiation field $rad$ serves as energy drain 
for the cavity mode, it also can be used to pump the mode.
Hamiltonian of the model is
\begin{equation}
 \label{H}
 \begin{split}
 \hat H &= \hat H_0 + \hat H_1 + \hat H_{vib}
 \\
 \hat H_0 &= \hat H_{sys} + \hat H_{bath}
 \\
 \hat H_1 &= \hat V_{ML} + \hat V_{MR} + \hat V_{c,rad}
\end{split}
\end{equation}
where
\begin{equation}
\label{Hsysbath}
\begin{split}
 \hat H_{sys} &= \hat H_M + \hat H_{c} + \hat V_{M,c}
 \\
 \hat H_{bath} &= \hat H_L + \hat H_R + \hat H_{rad}
\end{split}
\end{equation}
Here $\hat H_{sys}$ and $\hat H_{bath}$ are the system and baths Hamiltonians.
$\hat H_M$, $\hat H_{c}$, $\hat H_L$, $\hat H_R$ and $\hat H_{rad}$
are respectively Hamiltonians of the molecule, cavity mode, left and right contacts, and radiation field. 
Operators $\hat V$ introduce coupling between the subsystems.
$\hat H_{vib}$ is Hamiltonian representing vibrational degrees of freedom.
Explicit expressions are
\begin{align}
\label{HM}
 \hat H_M &= \sum_{m=1,2} \varepsilon_m\hat d_m^\dagger\hat d_m
 -t\left(\hat d_1^\dagger\hat d_2+\hat d_2^\dagger\hat d_1\right)
\\
 \hat H_{c} &= \omega_{c} \hat a_{c}^\dagger\hat a_{c}
 \\
\hat H_K &= \sum_{k\in K} \varepsilon_k \hat c_k^\dagger \hat c_k\quad (K=L,R)
 \\
 \hat H_{rad} &= \sum_\alpha \omega_\alpha \hat a_\alpha^\dagger \hat a_\alpha
 \\
 \hat V_{M,c} &= U_{c}\left(\hat a_{c}+\hat a_{c}^\dagger\right)
             \left(\hat d_1^\dagger\hat d_2+\hat d_2^\dagger\hat d_1\right)
 \\
 \hat V_{MK} &= \sum_{k\in K}\left( V_{mk}\hat d_m^\dagger\hat c_k + H.c.\right)
 \quad (m=1(2) \mbox{ for } K=L (R))
 \\
 \hat V_{c,rad} &= \sum_\alpha 
\left( U_\alpha \hat a_{c}^\dagger \hat a_\alpha + H.c. \right)
 \\
 \label{Hvib}
 \hat H_{vib} &= \frac{1}{2}\sum_{v=1,2} \left(\hat P_v^2+ \hat Q_v^2\right)
 + M^{(1)}\left(\hat d_1^\dagger\hat d_2+\hat d_2^\dagger\hat d_1\right) \hat Q_1
 + M^{(2)}\left(\hat d_1^\dagger\hat d_1-\hat d_2^\dagger\hat d_2\right)\hat Q_2
\end{align}
Here $\hat d_m^\dagger$ ($\hat d_m$) and $\hat c_k^\dagger$ ($\hat c_k$)
create (annihilate) electron on molecular level $m$ and state $k$ 
of the contacts, respectively. $\hat a_{c}^\dagger$ ($\hat a_{c}$)
and $\hat a_\alpha^\dagger$ ($\hat a_\alpha$) excites (destroys)
quanta in the cavity mode and mode $\alpha$ of the radiation field.
Note that Hamiltonian $\hat H_{vib}$ representing vibrational degrees of 
freedom and their coupling to electronic degrees of freedom is only used 
to derive expression for the electronic friction (see Ref.~\cite{chen_current-induced_2019} for details)
and does not participate in further numerical analysis.
Indeed, electron induced nuclear forces (including friction) can be introduced only for classical nuclei
Thus, vibrational Hamiltonian $\hat H_{vib}$ is only necessary as a starting point of a derivation
reducing full quantum description of nuclei to their classical behavior. 
 After the derivation is finished, expression for the forces (including friction) depend on electronic degrees of freedom only
 and vibrational Hamiltonian does not participate in further analysis. 
Still, we show the $\hat H_{vib}$ because it defines electron-nuclear coupling $M^{(1,2)}$ used in the considerations below.
Note that in realistic simulation for strictly adiabatic limit of nuclear dynamics 
electronic structure depends on static nuclear configuration. This means that
parameters of electronic Hamiltonian (such as level positions $\varepsilon_m$, electron hopping $t$, coupling to contacts 
$V_{mk}$, etc.) depend on nuclear positions. In the analysis below we utilize set of fixed parameters, which may be considered 
as values corresponding to one nuclear frame. 

We note that in (\ref{Hsysbath}) $\hat V_{M,c}$ was put into the system
Hamiltonian $H_{sys}$ to allow consideration of strong light-matter interaction. 
Clearly, quasiparticle representation is not the most convenient way
to treat $\hat V_{M,c}$. Instead, we utilize the Born-Oppenheimer type
many-body states as a basis for our consideration
\begin{equation}
 \lvert S\rangle = \lvert e\rangle\,\lvert p\rangle
\end{equation}
Here $\lvert e\rangle$ represents one of four possible electronic states:
empty state $\lvert 0,0\rangle$, electron in level $1$ $\lvert 1,0\rangle$, 
electron in level $2$ $\lvert 0,1\rangle$, and two electrons in the molecule
$\lvert 1,1\rangle$.
$\lvert p\rangle$ are states of the harmonic oscillator representing the 
cavity mode. Using spectral decompositions of the second quantized operators 
in the many-body states 
\begin{align}
\hat d_m^\dagger &= \sum_{S_1,S_2} \lvert S_2\rangle
 \langle S_2\rvert \hat d_m^\dagger \lvert S_1\rangle\langle S_1\rvert
 \equiv \sum_{S_1,S_2}\delta_{p_2,p_1}\xi^m_{e_2,e_1} \hat X_{S_1S_2}^\dagger
 \\
 \hat a_{c}^\dagger &= \sum_{S_1,S_2} \lvert S_2\rangle
 \langle S_2\rvert \hat a_{c}^\dagger \lvert S_1\rangle\langle S_1\rvert
 \equiv \sum_{S_1,S_2}\delta_{e_2,e_1} \delta_{p_2,p_1+1}\sqrt{p_2}\,
 \hat X_{S_1S_2}^\dagger
\end{align}
one can represent the model in many-body basis of the zero-order Hamiltonian
with baths ($L$, $R$, and $rad$) still represented in standard second
quantization (see Appendix~\ref{appA} for explicit form of the Hamiltonian).

After transformation to many-body eigenstates of the Hamiltonian $\hat H_{sys}$,
\begin{equation}
 \hat H_{sys} = \sum_S E_S \hat X_{SS},
\end{equation}
we introduce single-particle Hubbard Green's function
\begin{equation}
 \label{GF}
 G_{(S_1S_2)(S_3S_4)}(\tau,\tau') = -i\langle T_c\, \hat X_{S_1S_2}(\tau)\,
\hat X_{S_3S_4}^\dagger(\tau')\rangle
\end{equation}
Here $T_c$ is the Keldysh contour ordering operator, 
$\tau$ and $\tau'$ are the contour variables, and
\begin{equation}
 \label{Hub_op}
 \hat X_{S_1S_2} \equiv \lvert S_1\rangle\langle S_2\rvert
\end{equation}
is the Hubbard operator.
Following Ref.~\cite{ChenOchoaMGJCP17} one has to solve the modified Dyson equation
with self-energies due to coupling to the contacts and to the radiation 
field evaluated within nonequilibrium diagrammatic technique
for the Hubbard Green functions. 
The solution is self-consistent because the self-energies both define Green functions
(via the modified Dyson equation) and depend on them
(via self-energies expressions). 
In the consideration below we utilize second order diagrammatic expansion 
in the system-baths couplings. Short details about the self-consistent
procedure and explicit forms of the self-energies are given in 
Appendix~\ref{appB}. 

Once Green function (\ref{GF}) is known, following our derivation
in Ref.~\cite{chen_current-induced_2019} we calculate electronic friction for the model (\ref{H}) as
\begin{equation}
 \label{friction}
 \gamma_{ab}(E) = \sum_{S_1,S_2,S_3,S_4}
 M^{(a)}_{S1S_2}\, G^r_{(S_1S_2)(S_3S_4)}(E)\, M^{(b)}_{S_4S_3}
\end{equation}
where $M^{(a)}_{S_1S_2}$ and $M^{(b)}_{S_4S_3}$ ($a,b=1,2$)
are the electron-vibration interactions $M^{(1)}$ and $M^{(2)}$
introduced in (\ref{Hvib}) represented in the many-body eigenbasis
of the Hamiltonian $\hat H_0$.
$G^{r}_{(S_1S_2)(S_3S_4)(E)}$ is the Fourier transform of 
retarded projection of the Hubbard Green's function (\ref{GF}).
We note in passing that the friction tensor (\ref{friction}) has two nuclear indices because nuclear quantum deviations 
from classical trajectory in  derivation of Ref.~\cite{chen_current-induced_2019} are taken into account up to second order 
in cumulant expansion. Higher order expansion would result in more nuclear indices (one for every additional order).
We also note that while in the model (for simplicity and in order to demonstrate accuracy of our Hubbard NEGF method in
non-interacting case, where exact solution is known) we consider linear electron-nuclei coupling, 
the derivation in Ref.~\cite{chen_current-induced_2019} and expression for friction are more general.


\section{Results and discussion}\label{numres}
Unless stated otherwise parameters of the simulations are the following. Simulations are performed at room temperature $T=300$~K
and molecular levels are set as  $\varepsilon_1=-\varepsilon_2=0.1$~eV with electron hopping parameter $t=0.1$~eV.
Strength of coupling between electronic and vibrational degrees of freedom is $M^{(1)}=M^{(2)}=0.01$~eV 
and molecular exciton coupling to cavity mode (the strong light-matter interaction parameters) is $U_{c}=0.5$~eV.
Electron escape rates to metallic contacts
\begin{equation}
\Gamma^K_{m_1m_2}(E) \equiv 2\pi\sum_{k\in K} V_{m_1k} V_{km_2} \delta(E-\varepsilon_k)
\end{equation}
are assumed to be energy independent (the wide band approximation), they are
$\Gamma^{L(R)}_{m_1m_2}=\delta_{m_1,m_2}\,\delta_{m_1,1(2)}\, 0.1$~eV.
Frequency of the cavity mode is taken as $\omega_{c}=0.2$~eV and the mode energy dissipation rate
\begin{equation}
 \gamma_{c}(\omega) \equiv 2\pi\sum_{\alpha} \lvert U_\alpha\rvert^2 \delta(\omega-\omega_\alpha)
\end{equation}
(also assumed to be energy independent) is $0.01$~eV. 
Fermi energy is taken as an origin, $E_F=0$, and bias $V_{sd}$ is applied symmetrically, 
$\mu_L=E_F+\lvert e\rvert V_{sd}/2$ and $\mu_R=E_F-\lvert e\rvert V_{sd}/2$.
Unless specified otherwise, we take $V_{sd}=1$~V.
Radiation field is modeled as continuum of modes with modes around frequency of  the laser, $\omega_{0}$ being populated as
\begin{equation}
 N(\omega) = I_0\, \frac{\delta^2}{(\omega-\omega_0)^2+\delta^2}
\end{equation} 
Here $\delta$ is the laser bandwidth and $I_0$ is its intensity. Below we take $\omega_0=0.2$~eV
and $\delta=0.1$~eV. $I_0=1$ in the presence of pumping and $0$ otherwise.
Simulations are performed on an adjustable grid. The self-consistent calculation is assumed to be converged
when populations of many-body states at subsequent steps of the procedure differ no more than $0.01$.

\begin{figure}[htbp]
\includegraphics[width=\linewidth]{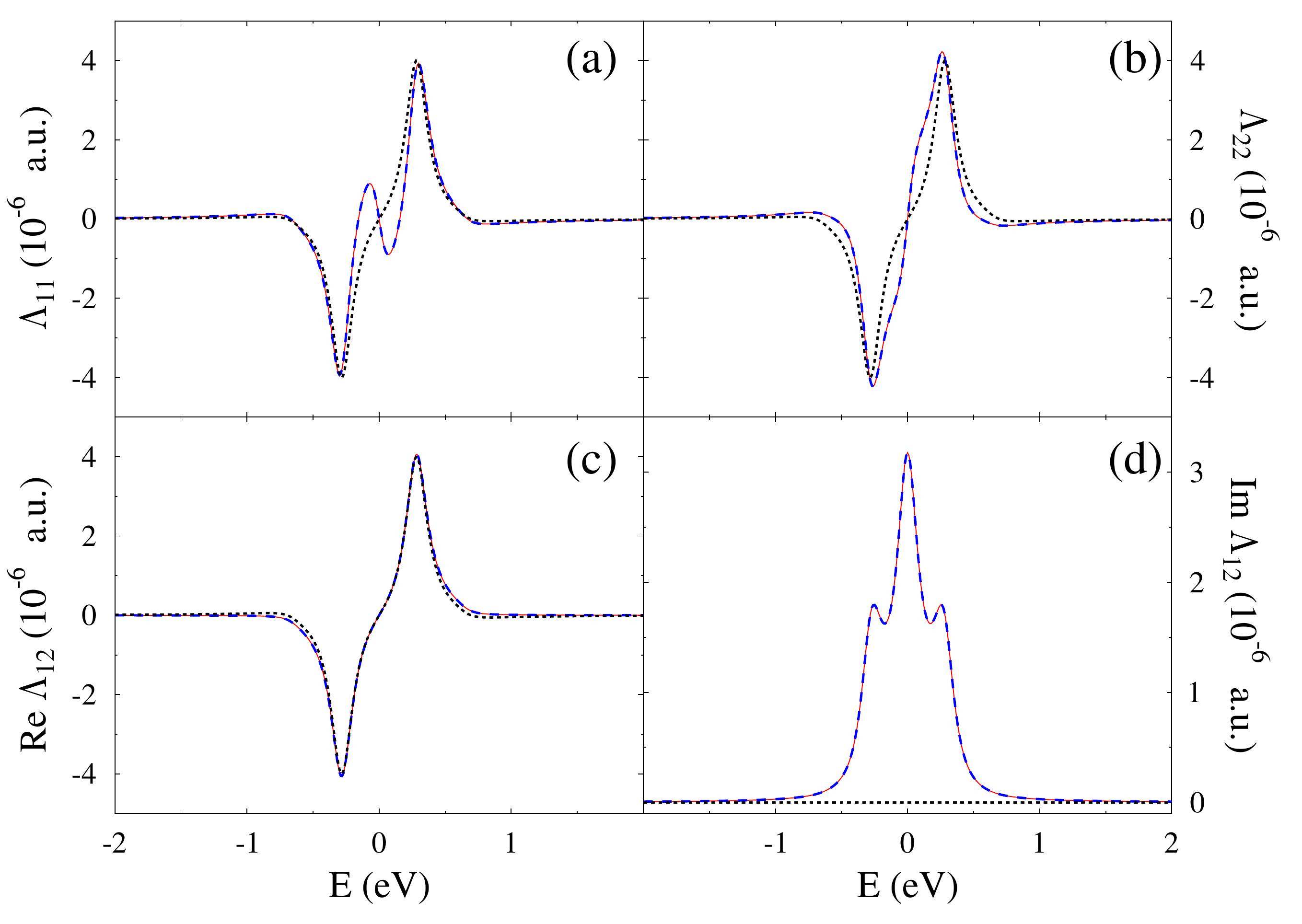}
\caption{\label{fig1}
Friction tensor $\Lambda(E)$, Eq.~(\ref{Lambda}), for non-interacting model, $U_{c}=0$. 
Shown are exact result (solid line, red), the Hubbard NEGF simulation (dashed line, blue),
and results obtained using nonequilibrium generalization of the Head-Gordon and Tully  electronic friction
(dotted line, black). 
Parameters of the simulations are $T=300$~K, $\varepsilon_1=-\varepsilon_2=0.1$~eV, $t=0.1$~eV,
$M^{(1)}=M^{(2)}=0.01$~eV, $\Gamma^{L(R)}_{m_1m_2}=\delta_{m_1,m_2}\,\delta_{m_1,1(2)}\, 0.1$~eV, and
$V_{sd}=1$~V.
}
\end{figure}

We start from a non-interacting case, $U_{c}=0$, exact results for which were originally derived in Ref.~\cite{TodorovPRB12}.
To facilitate comparison with that work, we consider the $\Lambda_{ab}(E)$ function, which is related to the friction tensor 
(\ref{friction}) in time domain as
\begin{equation}
\label{Lambda}
\gamma_{ab}(t) = 2\pi i\, \theta(t)\, \Lambda_{ab}(t)
\end{equation}
Figure~\ref{fig1} shows elements of the tensor simulated within the standard NEGF (exact result) and the Hubbard NEGF.
We also show results calculated using generalized version of the celebrated Head-Gordon and Tully (HGT) expression for 
electronic friction. 
Note that the original HGT expression is derived from consideration of  a time-dependent Schr{\" o}dinger equation
and states of electronic system are expressed in terms of `single determinantal wave functions.
Note also that the original HGT friction kernel is of the second order in non-adiabatic transfer element.
This means that the original consideration is performed for an isolated molecule (no baths at all), 
and that the consideration is restricted to non-interacting  (mean-filed) electronic systems
with weak electron-nuclei coupling. Our recent publication, Ref.~\cite{chen_current-induced_2019}, generalizes 
the HGT expression to open nonequilibrium interacting (beyond mean field) systems with arbitrary (both in form and strength)
electron-nuclei coupling (see Ref.~\cite{chen_current-induced_2019} for details).
However, even this generalized version of the HGT expression misses bath-induced coherences
which become important in quasi-degenerate situations when energy separation between many-body states
of electronic system are smaller than characteristic energy scale of the system-bath interaction.
Below we demonstrate failure of the expression comparing it to results of the Hubbard NEGF simulations.
Fig.~\ref{fig1} is similar to Fig.~3 of Ref.~\cite{chen_current-induced_2019}; but calculated for a different set of parameters.
As previously, Hubbard NEGF is pretty accurate in reproducing exact results. For smaller $\Gamma$ considered 
here, the generalized Head-Gordon and Tully expression becomes quite accurate near molecular resonances
($\sqrt{(\varepsilon_1-\varepsilon_2)^2+4t^2}$), while still missing coherence related contributions: 
$\Lambda_{11}(E)$ near $E=0$ and $\mbox{Im}\,\Lambda_{12}(E)$. 
We note in passing that at $E>0$ areas with $\Lambda<0$ correspond to usual friction (i.e. situation when electronic bath slows 
down nuclear motion), while areas with $\Lambda>0$ correspond to nuclear motion being speeded up by the electronic 
subsystem. Please note that such vibrational instability  (negative friction caused by a population-inverted situation) 
was discussed in prior publications (see, e.g., Ref.~\cite{TodorovPRB12}).
 
\begin{figure}[htbp]
\includegraphics[width=\linewidth]{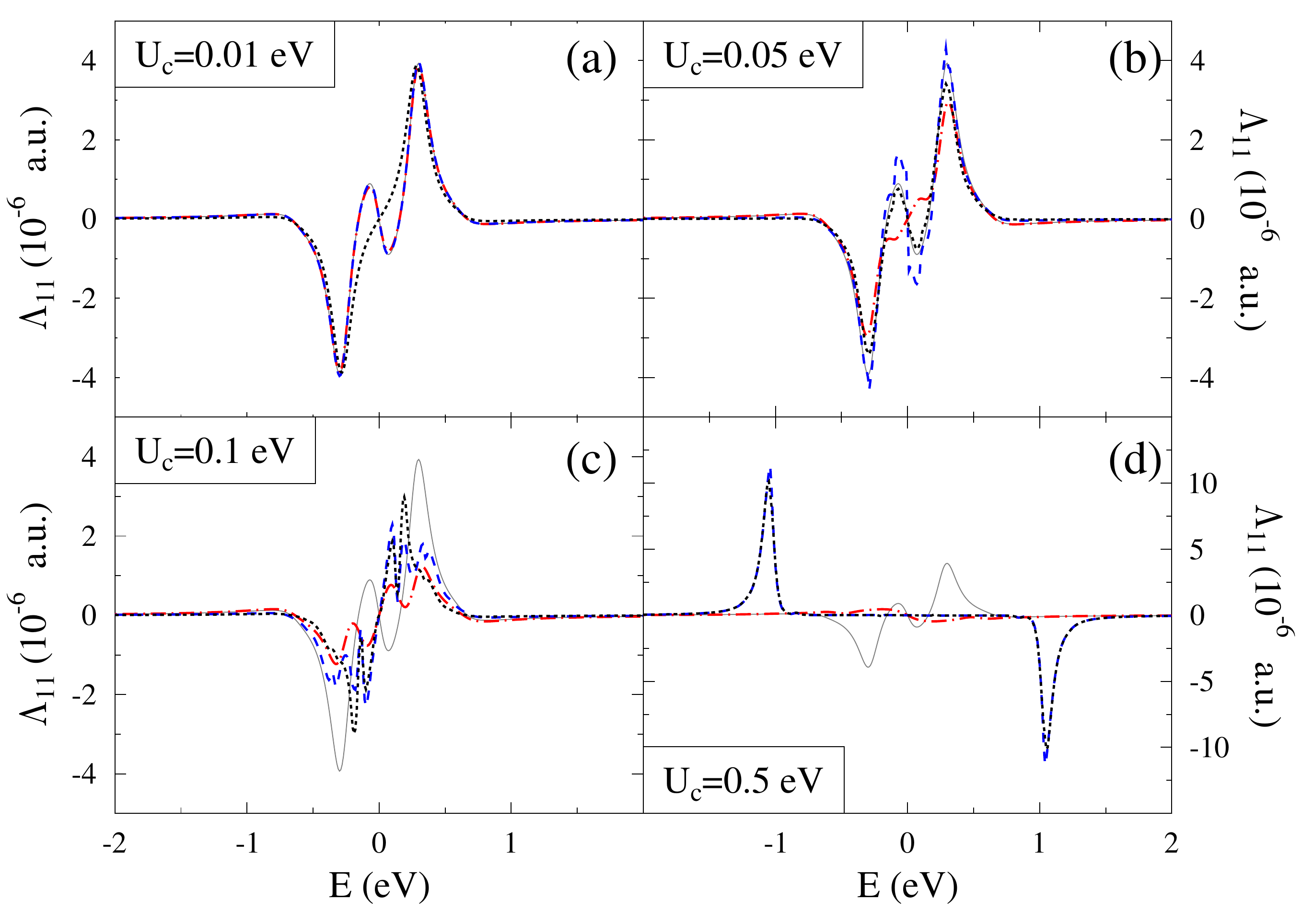}
\caption{\label{fig2}
Friction tensor $\Lambda_{11}(E)$, Eq.~(\ref{Lambda}), in the presence of molecular exciton-cavity mode coupling, $U_{c}$. 
Panels (a)-(d) show results for several values of the interaction.
Each panel presents results of simulations within the mean-field NEGF (dash-dotted line, red), the Hubbard NEGF (dashed line, blue),
and nonequilibrium generalization of the Head-Gordon and Tully  electronic friction (dotted line, black). 
For comparison, solid gray line shows exact non-interacting, $U_{c}=0$, result.  
Cavity mode frequency is $\omega_{c}=0.2$~eV. Other parameters are as in Fig.~\ref{fig1}. 
}
\end{figure}

Next we consider effect of molecular exciton-cavity mode coupling on electronic friction.
Figure~\ref{fig2} shows results of simulations for the $\Lambda_{11}$ elements of the friction tensor, Eq.~(\ref{Lambda}),
for a set of interaction strengths. We compare the Hubbard NEGF with the generalized version of the Head-Gordon and Tully 
expression and to mean-field NEGF treatment of Ref.~\cite{TodorovPRB12} (details of the NEGF Green function simulations
are given in Appendix~\ref{appC}). 
We see that for extremely weak interaction (Fig.~\ref{fig2}a) mean field and Hubbard NEGF yield the same result;
with the interaction growing (Fig.~\ref{fig2}b) the two approaches start to deviate from each other. 
For intermediate interaction strength (Fig.~\ref{fig2}c) molecule-cavity mode coupling slightly reduces electronic friction.
We attribute the effect to relatively lower values for the coupling $M^{(1)}$ for separate channels 
(transitions between different pairs of the many-body states) in the many-body eigenbasis.
At strong coupling (Fig.~\ref{fig2}d) the Hubbard NEGF shows enhanced electronic friction,
while mean field NEGF predicts reduction in the friction. 
The reason for discrepancy is importance of intra-system correlations between electronic and cavity mode degrees of freedom.
This is missed by the mean-field NEGF treatment. Note that inadequacy of mean-field predictions of electronic friction was
also shown in Refs.~\cite{SubotnikPRL17,hopjan_molecular_2018}.

The generalized Head-Gordon and Tully result quite expectedly becomes accurate in the case of strong coupling between
molecular exciton and cavity mode.
Indeed, strong intra-system interaction is equivalent to relatively weak system-bath coupling (i.e. we are in regime where bath
induced correlations between many-body eigenstates of the system become less important). 
Relative accuracy of the generalized Head-Gordon and Tully  expression in the case of weak system-bath interaction was 
demonstrated in our previous study~\cite{chen_current-induced_2019} for a non-interacting model.
As we showed in that study, nonequilibrium generalization of the Head-Gordon and Tully result comes from 
$S_1=S_3$ and $S_2=S_4$ subset of terms in the general expression (\ref{friction}).
For relatively weak system-bath coupling one can further simplify the analysis by going to quasiparticle limit~\footnote{Note that original Head-Gordon and Tully expression for electronic friction~\cite{TullyJCP95} comes from equilibrium quasiparticle consideration of the mentioned subset of terms.}. Under these approximations, electronic friction (\ref{Lambda}) can be written as
\begin{equation}
\label{Lambda_approx}
\Lambda_{ab}(E) \approx \sum_{S_1,S_2} M^{(a)}_{S_1S_2} M^{(b)}_{S_2S_1} \big(P_{S_2}-P_{S_1}\big)\,
\delta\big(E-\left[E_{S_2}-E_{S_1}\right]\big)
\end{equation}
Here $P_S$ is probability of the eigenstate $\lvert S\rangle$ to be populated and $E_S$ is its energy.

\begin{figure}[htbp]
\includegraphics[width=\linewidth]{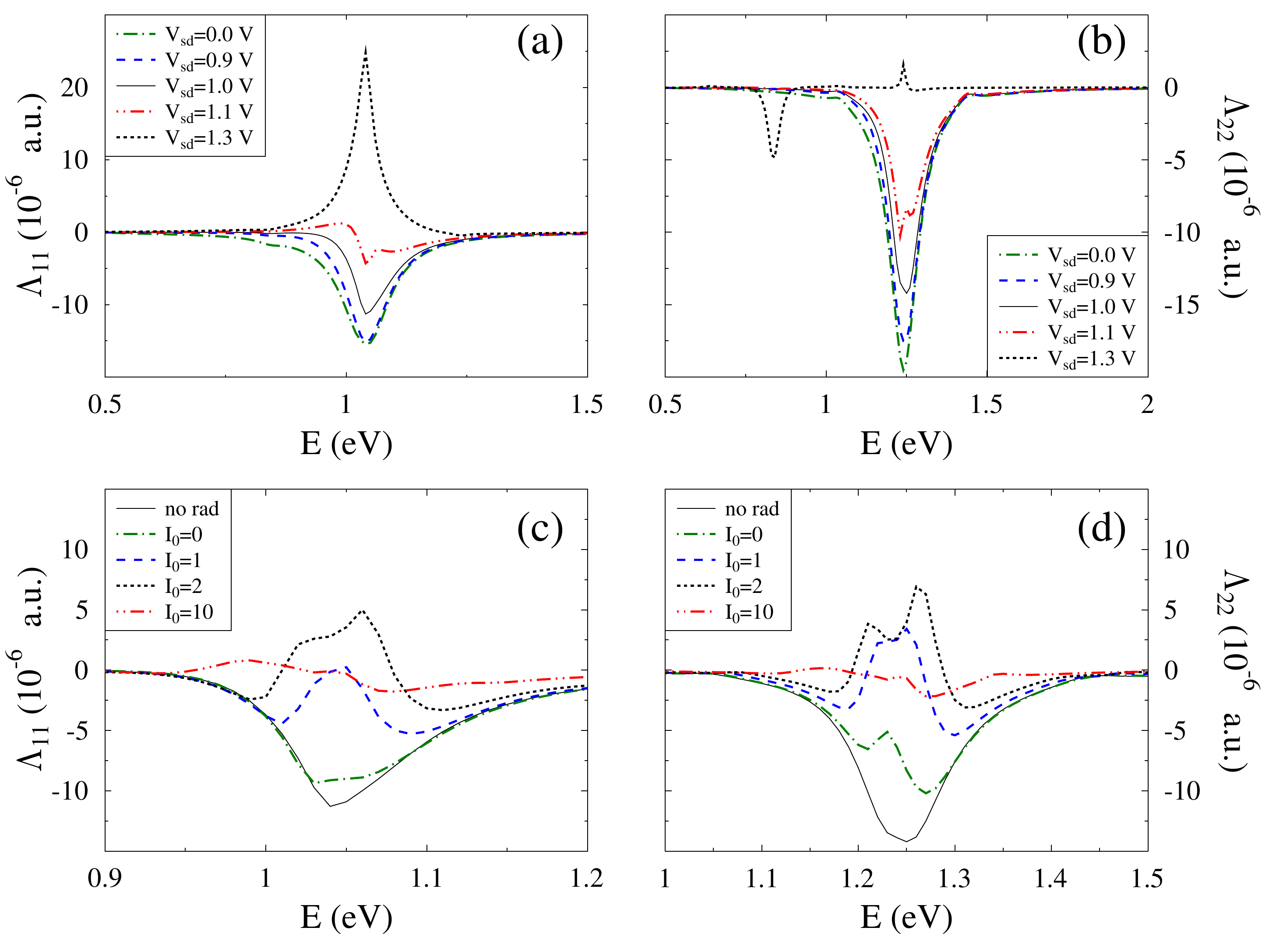}
\caption{\label{fig3}
Control of electronic friction by external perturbations in the strong coupling regime $U_{c}=0.5$~eV.
Shown are the Hubbard NEGF results for 
(a) $\Lambda_{11}(E)$ and (b) $\Lambda_{22}(E)$ for several biases $V_{sd}$ in the absence of radiation field;
(c) $\Lambda_{11}(E)$ and $\Lambda_{22}(E)$ for several intensities $I_0$ of radiation field at $V_{sd}=1$~V.
Pumping radiation field is taken at resonance with cavity mode, $\omega_0=0.2$~eV. Laser bandwidth is $\delta=0.1$~eV.
Simulations presented in panels (a) and (b) are performed in the absence of pumping, $I_0=0$; 
bias in simulations presented in panels (c) and (d) is $V_{sd}=1$~V. Other parameters are as in Fig.~\ref{fig2}.
}
\end{figure}

Eq.~(\ref{Lambda_approx}) indicates that peaks in the friction correspond to electronic transitions within charging block
and that sign and value of the friction are defined by probabilities of the corresponding many-body eigenstates.   
Thus, controlling the probabilities allows to engineer the friction: equal probabilities yield zero contribution of the transition
to electronic friction, maximum contribution is for transition between empty and filled states. 
Such control can be achieved, e.g., by changing external bias or by pumping the cavity mode.
Figures~\ref{fig3}a and b show control of electronic friction for vibrational modes $1$ and $2$, respectively.
Simulations are performed within the Hubbard NEGF for for strong light-matter coupling, $U_{c}=0.5$~eV.
One sees that friction dependence on the bias is non-monotonic. That is, at higher biases and hence for higher currents, 
one can get smaller friction. For the considered parameters $V_{sd}=1.1$~V minimizes friction for the mode $1$
and $V_{sd}=1.3$~V friction is minimal for mode $2$.
Similarly, control of electronic friction for the two vibrational modes by pumping cavity mode with external radiation field 
is shown in Figures~\ref{fig3}c and d, respectively. Note that even in absence of pumping, $I_0=0$, friction for 
isolated cavity mode differs from the result for mode coupled to empty radiation field due to energy damping in the latter case.
Also here friction behaves non-monotonically with intensity of the field, and for the parameters chosen minimum
of electronic friction is achieved for $I_0=10$.
 
\begin{figure}[htbp]
\includegraphics[width=\linewidth]{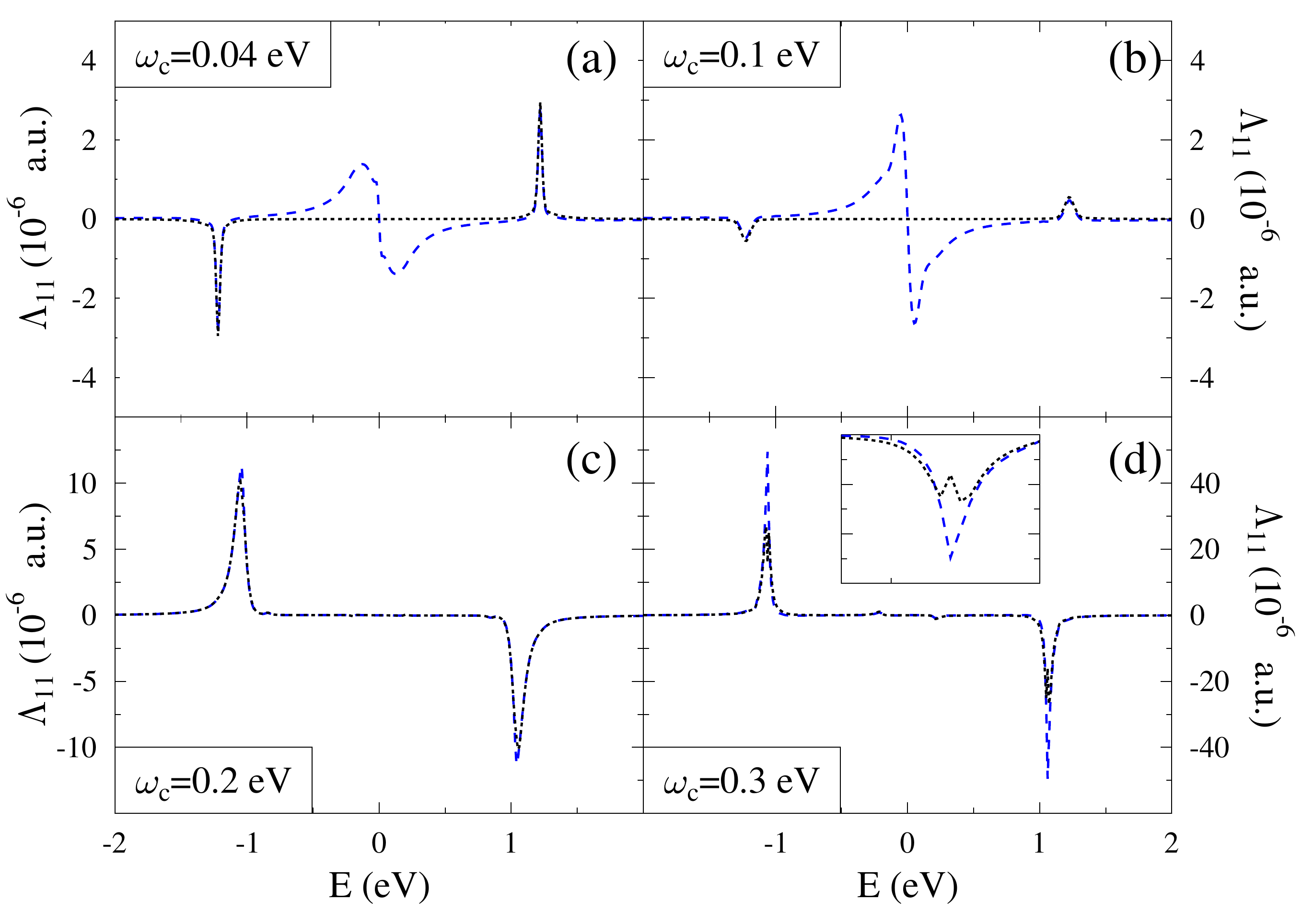}
\caption{\label{fig4}
Friction tensor $\Lambda_{11}(E)$, Eq.~(\ref{Lambda}), in the presence of molecular exciton-cavity mode coupling, $U_{c}=0.5$~eV.
Panels (a)-(d) show results for several values of cavity mode frequency $\omega_{c}$. 
Each panel presents results of simulations within the Hubbard NEGF (dashed line, blue),
and nonequilibrium generalization of the Head-Gordon and Tully  electronic friction (dotted line, black). 
Inset in panel (d) shows the peak at $E\sim 1$~eV in more details. 
Simulations are performed at bias $V_{sd}=1$~V in the absence of pumping, $I_0=0$.
Other parameters are as in Fig.~\ref{fig3}.
}
\end{figure}

Finally, we note that  even at strong light-matter interaction, the generalized Head-Gordon and Tully
expression is accurate only near resonance, $\omega_{c}\approx \varepsilon_1-\varepsilon_2$.
In an off-resonant situation bath-induces coherence between close (relative to $\Gamma$) molecular resonances,
are beyond the expression capabilities. Figure~\ref{fig4} shows electronic friction $\Lambda_{11}(E)$
at strong exciton-cavity mode coupling of $U_{c}=0.5$~eV for a set of cavity mode frequencies.
One sees that  the Hubbard NEGF coincides with the Head-Gordon and Tully result only at resonant $\omega_{c}$
(see Fig.~\ref{fig4}c).
Any shift out of the resonance condition, so that $\omega_{c}-\lvert \varepsilon_1-\varepsilon_2\rvert\geq\Gamma$,
leads to discrepancy between the two results (see Figs.~\ref{fig4}a, b, and d).


\section{Conclusion}\label{conclude}
We discuss electronic friction in interacting systems. In particular, we consider a model of a junction consisting of 
single molecule in nano cavity with strong light-matter interaction. 
Such systems have been realized experimentally; however, so far measurements were restricted to unbiased junctions.
Recently, we derived general expression for electronic friction in interacting systems~\cite{chen_current-induced_2019}. 
We showed that the friction can be conveniently expressed in terms of retarded projection of 
the single-particle Hubbard Green function. Here, we use introduce by us nonequilibrium diagrammatic technique
for the Hubbard NEGF~\cite{ChenOchoaMGJCP17} to calculate electronic friction in the interacting system.

We compare the Hubbard NEGF wit the mean-field standard NEGF treatment of Ref.~\cite{TodorovPRB12}
and with the nonequilibirum generalization~\cite{chen_current-induced_2019} of the Head-Gordon and Tully 
expression~\cite{TullyJCP95}. As expected standard and Hubbard NEGF results coincide in non- and weakly interacting
systems, while for strong intra-system interactions (coupling between molecular exciton and cavity mode)
the mean-field treatment fails qualitatively. This observation is in agreement with
previous studies in Refs.~\cite{SubotnikPRL17,hopjan_molecular_2018}. 

The generalized expression for the friction is shown to be quite accurate at strong
coupling resonant conditions, where energetic separation of electronic transitions between many-body states of 
the system is larger than strength of system-baths couplings. 
However, the expression misses bath-induced coherences between the transitions,
so that for small coupling or in off-resonant situation treatment beyond the Head-Gordon and Tully result is required.

A simple qualitative analysis based on approximate quasiparticle limit of the general expression 
shows that electronic friction depends on probabilities of pairs of many-body states involved in electronic transition:
friction is maximum for bi difference in the probabilities and approaches zero for equally probable states.
The latter can be modified with external perturbations such as bias and optical pumping. 
The analysis is confirmed with the Hubbard NEGF simulations.

Further development of the Hubbard NEGF method, generalization of the study to realistic systems,
combining The Hubbard NEGF with {\em ab initio} simulations and exploring possibilities to control molecular dynamics in 
nano-cavities are the goals for future research.

\begin{acknowledgments}
MG research is supported by the National Science Foundation (CHE-1565939)
and the US Department of Energy (DE-SC0018201).
\end{acknowledgments}

\appendix

\section{Molecular many-body basis representation of the model}\label{appA}
In the Born-Oppenheimer like basis of many-body states $\lvert S\rangle=\lvert e\rangle\,\lvert p\rangle$
in the molecular subspace the model (\ref{H})-(\ref{Hvib}) takes the form
\begin{equation}
\begin{split}
& \hat H_M + \hat H_{c} + \hat V_{M,c}  = \sum_{S_1,S_2} H_{S_1S_2} \hat X_{S_1S_2}
\\
&  \hat V_{MK} = \sum_{S_1,S_2}\sum_{k\in K}\left( V_{(S_1S_2)k}\hat X_{S_1S_2}^\dagger\hat c_k + H.c.\right)
 \\
& \hat V_{c,rad} = \sum_{S_1,S_2}\sum_\alpha \left(U_{(S_1S_2)\alpha}\hat X_{S_1S_2}^\dagger\hat a_\alpha + H.c.\right)
 \\
& \hat H_{vib} = \frac{1}{2}\sum_{v=1,2} \left(\hat P_v^2+ \hat Q_v^2\right)
 + \sum_{S_1,S_2}\sum_{v=1,2}M^{(v)}_{S_1S_2}\hat X_{S_1S_2} \hat Q_v
 \end{split}
\end{equation}
where
\begin{equation}
\begin{split}
 H_{S_1S_2} &= \delta_{p_1,p_2}\left(\sum_S\sum_{m=1,2}\varepsilon_m\xi^m_{S_1S}\overset{*}{\xi}{}^{m}_{S_2S}
 - t \sum_S\left( \xi^1_{S_1S}\overset{*}{\xi}{}^{2}_{S_2S} + c.c.\right) \right)
 \\ &
 +\delta_{e_1,e_2}\delta_{p_1,p_2}\omega_{c}\, p_1
 + U_{c}\left(\delta_{p_1+1,p_2}\sqrt{p_2}+\delta_{p_1,p_2+1}\sqrt{p_1}\right)\sum_S\left(\xi^{1}_{S_1S}\overset{*}{\xi}{}^{2}_{S_2S}+c.c.\right)
 \\
 V_{(S_1S_2)k} &= \sum_m V_{mk} \xi^m_{S_2S_1}
 \\
 U_{(S_1S_2)\alpha} &= U_\alpha\, \chi^{c}_{S_2S_1}
 \\
 M^{(1)}_{S_1S_2} &= M^{(1)} \sum_S\left(\xi^{1}_{S_1S}\overset{*}{\xi}{}^{2}_{S_2S}+c.c.\right)
 \\
 M^{(2)}_{S_1S_2} &= M^{(2)} \sum_S\sum_{m=1,2} (-1)^{m+1}\xi^m_{S_1S}\overset{*}{\xi}{}^{m}_{S_2S}
 \end{split}
\end{equation}
 $H_K$ ($K=L,R$) and $H_{rad}$ are kept in second quantized form.
 We transfer to eigenbasis of the $\hat H_M + \hat H_{c} + \hat V_{M,c}$ Hamiltonian before
 starting the Hubbard NEGF simulations. 


\section{The Hubbard NEGF simulations}\label{appB}
Details of the Hubbard NEGF method can be found in Ref.~\cite{ChenOchoaMGJCP17}.  
Here we give short summary of the procedure utilized in the simulations.
In the model system (electronic and cavity mode degrees of freedom) are coupled to three baths:
two Fermi baths (contacts $L$ and $R$) and one Boson bath (radiation field).
The Hubbard NEGF is a diagrammatic technique expanding in system-baths coupling strengths.
We work in the lowest (second) order of the expansion. Moreover, because main contribution comes from 
single electron transfer events (transitions between molecule and contacts and intra-system excitations),
for simplicity we restrict our consideration to first Hubbard approximation. 

Within the approach, one has to solve Dyson equation for locators
\begin{equation}
 \big(i\partial_{\tau_1}-\Delta_{\mathcal{M}_1}\big)g_{\mathcal{M}_1\mathcal{M}_2}(\tau_1,\tau_2)
 -\sum_{\mathcal{M}}\int_cd\tau\,\Sigma_{\mathcal{M}_1\mathcal{M}}(\tau_1,\tau)
 g_{\mathcal{M}\mathcal{M}_2}(\tau,\tau_2) = 
 \delta_{\mathcal{M}_1,\mathcal{M}_2}\delta(\tau_1,\tau_2)
\end{equation}
from which Hubbard Green's function is obtained by multiplication with spectral weight
$P_{\mathcal{M}_1\mathcal{M}_2}$
\begin{equation}
G_{\mathcal{M}_1\mathcal{M}_2}(\tau_1,\tau_2) = \sum_\mathcal{M}
g_{\mathcal{M}_1\mathcal{M}}(\tau_1,\tau_2)\, P_{\mathcal{M}\mathcal{M}_2}(\tau_2)
\end{equation} 
Here  $\mathcal{M}$ is Fermi type transition (electron transition between two many-body states, which differ by one electron),
\begin{equation}
P_{\mathcal{M}_1\mathcal{M}_2}(\tau)\equiv \bigg\langle
\bigg\{\hat X_{\mathcal{M}_1}(\tau);\hat X_{\mathcal{M}_2}^\dagger(\tau)\bigg\}\bigg\rangle
\end{equation}
and $\Sigma_{\mathcal{M}_1\mathcal{M}_2}(\tau_1,\tau_2)$
is Hubbard self-energy, which consists from contributions of self-energies
due to coupling to contacts ($K=L,R$) and radiation field ($rad$).

Explicit expressions for the self-energies are
\begin{equation}
\begin{split}
\Sigma_{\mathcal{M}_1\mathcal{M}_2}(\tau_1,\tau_2) &= \sum_{K=L,R}\sum_{\mathcal{M}_3} P_{\mathcal{M}_1\mathcal{M}_3}(\tau_1)\,
\sigma^K_{\mathcal{M}_3\mathcal{M}_2}(\tau_1,\tau_2)
\\
&+ i\sum_{\mathcal{B}_1,\mathcal{B}_2} \big( 
 s_3(\mathcal{M}_1,\mathcal{B}_1)\,\pi_{\mathcal{B}_1\mathcal{B}_2}(\tau_1,\tau_2)\,s_4(\mathcal{M}_3,\tilde{\mathcal{B}}_2)
 \\ &\qquad\quad
+ s_3(\mathcal{M}_1,\tilde{\mathcal{B}}_1)\,\pi_{\mathcal{B}_2\mathcal{B}_1}(\tau_2,\tau_1)\,s_4(\mathcal{M}_3,\mathcal{B}_2)
\big) g_{\mathcal{M}_3\mathcal{M}_2}(\tau_1,\tau_2)
\end{split}
\end{equation}
where $\mathcal{B}\equiv(S_1,S_2)$ is Bose type transition in the same charging block (electron transition between two 
many-body states with the same number of electrons), $\tilde{\mathcal{B}}=(S_2,S_1)$, 
\begin{equation}
s_3(\mathcal{M},\mathcal{B}) \mathcal{M}_3 = \delta_{2\mathcal{M},2\mathcal{B}}\,\hat X_{1\mathcal{M},1\mathcal{B}}
-\delta_{1\mathcal{M},1\mathcal{B}}\,\hat X_{2\mathcal{B},2\mathcal{M}}
\end{equation}
with $n\mathcal{M}$ and $n\mathcal{B}$ ($n=1,2$) being $n^{th}$ many-body state in the transitions, and
$\sigma$ and $\pi$ are usual NEGF self-energies due to coupling to respectively Fermi and Bose baths
\begin{align}
& \sigma^K_{\mathcal{M}_1\mathcal{M}_2}(\tau_1,\tau_2) = \sum_{k\in K} V_{\mathcal{M}_1k}\, g_k(\tau_1,\tau_2)\,
V_{k\mathcal{M}_2}
\\
& \pi_{\mathcal{B}_1\mathcal{B}_2}(\tau_1,\tau_2) = \sum_{\alpha} U_{\mathcal{B}_1\alpha}\, f_\alpha(\tau_1,\tau_2)\,
U_{\alpha\mathcal{B}_2}
\end{align}
Here 
\begin{align}
 g_k(\tau_1,\tau_2) &= -i \langle T_c\,\hat c_k(\tau_1\,\hat c_k^\dagger(\tau_2)\rangle
 \\
 f_\alpha(\tau_1,\tau_2) &= -i\langle T_c\, \hat a_\alpha(\tau_1)\,\hat a_\alpha^\dagger(\tau_2)\rangle
\end{align}


\section{Mean field NEGF simulations}\label{appC}
Within the NEGF we treat electron-cavity mode coupling at the SCBA level. 
Electron $G$ and cavity mode $F$ Green functions 
\begin{align}
 G_{m_1m_2}(\tau_1,\tau_2) &= -i\langle T_c\, \hat d_{m_1}(\tau_1)\,\hat d_{m_2}^\dagger(\tau_2)\rangle
 \\
 F_{c}(\tau_1,\tau_2) &= -i\langle T_c\, \hat a_{c}(\tau_1)\, \hat a^\dagger_{c}(\tau_2)\rangle
\end{align}
are solved utilizing the standard Dyson equation
\begin{align}
& \sum_m\left(\delta_{m_1,m}i\partial_{\tau_1}-H_{m_1m}\right) G_{mm_2}(\tau_1,\tau_2) =
   \sum_m \int_c d\tau\, \Sigma_{m_1m}(\tau_1,\tau)\, G_{mm_2}(\tau,\tau_2)
\\
& \left(i\partial_{\tau_1}-\omega_{c}\right) F_{c}(\tau_1,\tau_2) = \int_c d\tau\, \Pi(\tau_1,\tau)\, F_{c}(\tau,\tau_2)
\end{align}
with electron self-energy $\Sigma$ consisting of contributions due to coupling to the contacts and to cavity mode,
$\Sigma=\Sigma^L+\Sigma^R+\Sigma^{c}$,
and with cavity mode self-energy $\Pi$ consisting of contributions due to coupling to radiation field and electrons,
$\Pi=\Pi^{rad}+\Pi^{e}$.
Expressions for self-energies due to coupling to the baths have standard Fermi (contacts) and Bose (radiation filed) forms.
Electron-cavity mode coupling is treated at the second order of diagrammatic expansion so that
\begin{align}
\Sigma^{c}_{m_1m_2}(\tau_1,\tau_2) =& i \sum_{m_3,m_4} G_{m_3m_4}(\tau_1,\tau_2)
 \bigg(
 U^{c}_{m_1m_3}F_{c}(\tau_1,\tau_2)U^{c}_{m_4m_2}
 +U^{c}_{m_2m_4}F_{c}(\tau',\tau)U^{c}_{m_3m_1}
 \bigg)
 \\
  \Pi^{e}(\tau_1,\tau_2) =& 
-i \sum_{\begin{subarray}{c}m_1,m_2 \\ m_3,m_4\end{subarray}}
U^{c}_{m_1m_2}\, G_{m_2m_4}(\tau_1,\tau_2)\,
G_{m_3m_1}(\tau_2,\tau_1)\, U^{c}_{m_3m_4}
\end{align}
where $U^{c}_{m_1m_2}=\delta_{m_1,1}\delta_{m_2,2}+\delta_{m_1,2}\delta_{m_2,1}$.
For simplicity, in the simulations we utilize the quasiparticle limit for the phonon Green function.


\begin{thebibliography}{46}%
\makeatletter
\providecommand \@ifxundefined [1]{%
 \@ifx{#1\undefined}
}%
\providecommand \@ifnum [1]{%
 \ifnum #1\expandafter \@firstoftwo
 \else \expandafter \@secondoftwo
 \fi
}%
\providecommand \@ifx [1]{%
 \ifx #1\expandafter \@firstoftwo
 \else \expandafter \@secondoftwo
 \fi
}%
\providecommand \natexlab [1]{#1}%
\providecommand \enquote  [1]{``#1''}%
\providecommand \bibnamefont  [1]{#1}%
\providecommand \bibfnamefont [1]{#1}%
\providecommand \citenamefont [1]{#1}%
\providecommand \href@noop [0]{\@secondoftwo}%
\providecommand \href [0]{\begingroup \@sanitize@url \@href}%
\providecommand \@href[1]{\@@startlink{#1}\@@href}%
\providecommand \@@href[1]{\endgroup#1\@@endlink}%
\providecommand \@sanitize@url [0]{\catcode `\\12\catcode `\$12\catcode
  `\&12\catcode `\#12\catcode `\^12\catcode `\_12\catcode `\%12\relax}%
\providecommand \@@startlink[1]{}%
\providecommand \@@endlink[0]{}%
\providecommand \url  [0]{\begingroup\@sanitize@url \@url }%
\providecommand \@url [1]{\endgroup\@href {#1}{\urlprefix }}%
\providecommand \urlprefix  [0]{URL }%
\providecommand \Eprint [0]{\href }%
\providecommand \doibase [0]{http://dx.doi.org/}%
\providecommand \selectlanguage [0]{\@gobble}%
\providecommand \bibinfo  [0]{\@secondoftwo}%
\providecommand \bibfield  [0]{\@secondoftwo}%
\providecommand \translation [1]{[#1]}%
\providecommand \BibitemOpen [0]{}%
\providecommand \bibitemStop [0]{}%
\providecommand \bibitemNoStop [0]{.\EOS\space}%
\providecommand \EOS [0]{\spacefactor3000\relax}%
\providecommand \BibitemShut  [1]{\csname bibitem#1\endcsname}%
\let\auto@bib@innerbib\@empty
\bibitem [{\citenamefont {Bennett}\ \emph {et~al.}(2016)\citenamefont
  {Bennett}, \citenamefont {Kowalewski},\ and\ \citenamefont
  {Mukamel}}]{MukamelJCTC16}%
  \BibitemOpen
  \bibfield  {author} {\bibinfo {author} {\bibfnamefont {K.}~\bibnamefont
  {Bennett}}, \bibinfo {author} {\bibfnamefont {M.}~\bibnamefont {Kowalewski}},
  \ and\ \bibinfo {author} {\bibfnamefont {S.}~\bibnamefont {Mukamel}},\ }\href
  {\doibase 10.1021/acs.jctc.5b00824} {\bibfield  {journal} {\bibinfo
  {journal} {J. Chem. Theory Comput.}\ }\textbf {\bibinfo {volume} {12}},\
  \bibinfo {pages} {740} (\bibinfo {year} {2016})}\BibitemShut {NoStop}%
\bibitem [{\citenamefont {Petit}\ and\ \citenamefont
  {Subotnik}(2015)}]{SubotnikJCTC15}%
  \BibitemOpen
  \bibfield  {author} {\bibinfo {author} {\bibfnamefont {A.~S.}\ \bibnamefont
  {Petit}}\ and\ \bibinfo {author} {\bibfnamefont {J.~E.}\ \bibnamefont
  {Subotnik}},\ }\href {\doibase 10.1021/acs.jctc.5b00510} {\bibfield
  {journal} {\bibinfo  {journal} {J. Chem. Theory Comput.}\ }\textbf {\bibinfo
  {volume} {11}},\ \bibinfo {pages} {4328} (\bibinfo {year}
  {2015})}\BibitemShut {NoStop}%
\bibitem [{\citenamefont {Petit}\ and\ \citenamefont
  {Subotnik}(2014{\natexlab{a}})}]{SubotnikJCP14}%
  \BibitemOpen
  \bibfield  {author} {\bibinfo {author} {\bibfnamefont {A.~S.}\ \bibnamefont
  {Petit}}\ and\ \bibinfo {author} {\bibfnamefont {J.~E.}\ \bibnamefont
  {Subotnik}},\ }\href@noop {} {\bibfield  {journal} {\bibinfo  {journal} {J.
  Chem. Phys.}\ }\textbf {\bibinfo {volume} {141}},\ \bibinfo {eid} {154108}
  (\bibinfo {year} {2014}{\natexlab{a}})}\BibitemShut {NoStop}%
\bibitem [{\citenamefont {Petit}\ and\ \citenamefont
  {Subotnik}(2014{\natexlab{b}})}]{PetitSubotnikJCP14}%
  \BibitemOpen
  \bibfield  {author} {\bibinfo {author} {\bibfnamefont {A.~S.}\ \bibnamefont
  {Petit}}\ and\ \bibinfo {author} {\bibfnamefont {J.~E.}\ \bibnamefont
  {Subotnik}},\ }\href@noop {} {\bibfield  {journal} {\bibinfo  {journal} {J.
  Chem. Phys.}\ }\textbf {\bibinfo {volume} {141}},\ \bibinfo {eid} {014107}
  (\bibinfo {year} {2014}{\natexlab{b}})}\BibitemShut {NoStop}%
\bibitem [{\citenamefont {Jiang}\ \emph {et~al.}(2016)\citenamefont {Jiang},
  \citenamefont {Alducin},\ and\ \citenamefont {Guo}}]{GuoJPCL16}%
  \BibitemOpen
  \bibfield  {author} {\bibinfo {author} {\bibfnamefont {B.}~\bibnamefont
  {Jiang}}, \bibinfo {author} {\bibfnamefont {M.}~\bibnamefont {Alducin}}, \
  and\ \bibinfo {author} {\bibfnamefont {H.}~\bibnamefont {Guo}},\ }\href
  {\doibase 10.1021/acs.jpclett.5b02737} {\bibfield  {journal} {\bibinfo
  {journal} {J. Phys. Chem. Lett.}\ }\textbf {\bibinfo {volume} {7}},\ \bibinfo
  {pages} {327} (\bibinfo {year} {2016})}\BibitemShut {NoStop}%
\bibitem [{\citenamefont {Ouyang}\ \emph {et~al.}(2015)\citenamefont {Ouyang},
  \citenamefont {Saven},\ and\ \citenamefont {Subotnik}}]{SubotnikJPCC15}%
  \BibitemOpen
  \bibfield  {author} {\bibinfo {author} {\bibfnamefont {W.}~\bibnamefont
  {Ouyang}}, \bibinfo {author} {\bibfnamefont {J.~G.}\ \bibnamefont {Saven}}, \
  and\ \bibinfo {author} {\bibfnamefont {J.~E.}\ \bibnamefont {Subotnik}},\
  }\href {\doibase 10.1021/acs.jpcc.5b06655} {\bibfield  {journal} {\bibinfo
  {journal} {J. Phys. Chem. C}\ }\textbf {\bibinfo {volume} {119}},\ \bibinfo
  {pages} {20833} (\bibinfo {year} {2015})}\BibitemShut {NoStop}%
\bibitem [{\citenamefont {Lei}\ \emph {et~al.}(2016)\citenamefont {Lei},
  \citenamefont {Wu}, \citenamefont {Zheng}, \citenamefont {Zhai},\ and\
  \citenamefont {Zhu}}]{LeiZhuJPhotoChemPhotoBiolA16}%
  \BibitemOpen
  \bibfield  {author} {\bibinfo {author} {\bibfnamefont {Y.}~\bibnamefont
  {Lei}}, \bibinfo {author} {\bibfnamefont {H.}~\bibnamefont {Wu}}, \bibinfo
  {author} {\bibfnamefont {X.}~\bibnamefont {Zheng}}, \bibinfo {author}
  {\bibfnamefont {G.}~\bibnamefont {Zhai}}, \ and\ \bibinfo {author}
  {\bibfnamefont {C.}~\bibnamefont {Zhu}},\ }\href {\doibase
  10.1016/j.jphotochem.2015.10.025} {\bibfield  {journal} {\bibinfo  {journal}
  {J. Photochem. Photobiol. A}\ }\textbf {\bibinfo {volume} {317}},\ \bibinfo
  {pages} {39 } (\bibinfo {year} {2016})}\BibitemShut {NoStop}%
\bibitem [{\citenamefont {Wang}\ \emph {et~al.}(2015)\citenamefont {Wang},
  \citenamefont {Prezhdo},\ and\ \citenamefont
  {Beljonne}}]{PrezhdoBeljonnePCCP15}%
  \BibitemOpen
  \bibfield  {author} {\bibinfo {author} {\bibfnamefont {L.}~\bibnamefont
  {Wang}}, \bibinfo {author} {\bibfnamefont {O.~V.}\ \bibnamefont {Prezhdo}}, \
  and\ \bibinfo {author} {\bibfnamefont {D.}~\bibnamefont {Beljonne}},\ }\href
  {\doibase 10.1039/C5CP00485C} {\bibfield  {journal} {\bibinfo  {journal}
  {Phys. Chem. Chem. Phys.}\ }\textbf {\bibinfo {volume} {17}},\ \bibinfo
  {pages} {12395} (\bibinfo {year} {2015})}\BibitemShut {NoStop}%
\bibitem [{\citenamefont {Imada}\ \emph {et~al.}(2017)\citenamefont {Imada},
  \citenamefont {Miwa}, \citenamefont {Imai-Imada}, \citenamefont {Kawahara},
  \citenamefont {Kimura},\ and\ \citenamefont {Kim}}]{Imada2017}%
  \BibitemOpen
  \bibfield  {author} {\bibinfo {author} {\bibfnamefont {H.}~\bibnamefont
  {Imada}}, \bibinfo {author} {\bibfnamefont {K.}~\bibnamefont {Miwa}},
  \bibinfo {author} {\bibfnamefont {M.}~\bibnamefont {Imai-Imada}}, \bibinfo
  {author} {\bibfnamefont {S.}~\bibnamefont {Kawahara}}, \bibinfo {author}
  {\bibfnamefont {K.}~\bibnamefont {Kimura}}, \ and\ \bibinfo {author}
  {\bibfnamefont {Y.}~\bibnamefont {Kim}},\ }\href {\doibase
  10.1103/PhysRevLett.119.013901} {\bibfield  {journal} {\bibinfo  {journal}
  {Phys. Rev. Lett.}\ }\textbf {\bibinfo {volume} {119}},\ \bibinfo {pages}
  {013901} (\bibinfo {year} {2017})}\BibitemShut {NoStop}%
\bibitem [{\citenamefont {Imada}\ \emph {et~al.}(2016)\citenamefont {Imada},
  \citenamefont {Miwa}, \citenamefont {Imai-Imada}, \citenamefont {Kawahara},
  \citenamefont {Kimura},\ and\ \citenamefont {Kim}}]{Imada2016}%
  \BibitemOpen
  \bibfield  {author} {\bibinfo {author} {\bibfnamefont {H.}~\bibnamefont
  {Imada}}, \bibinfo {author} {\bibfnamefont {K.}~\bibnamefont {Miwa}},
  \bibinfo {author} {\bibfnamefont {M.}~\bibnamefont {Imai-Imada}}, \bibinfo
  {author} {\bibfnamefont {S.}~\bibnamefont {Kawahara}}, \bibinfo {author}
  {\bibfnamefont {K.}~\bibnamefont {Kimura}}, \ and\ \bibinfo {author}
  {\bibfnamefont {Y.}~\bibnamefont {Kim}},\ }\href {\doibase
  10.1038/nature19765} {\bibfield  {journal} {\bibinfo  {journal} {Nature}\
  }\textbf {\bibinfo {volume} {538}},\ \bibinfo {pages} {364} (\bibinfo {year}
  {2016})}\BibitemShut {NoStop}%
\bibitem [{\citenamefont {Lee}\ \emph {et~al.}(2016)\citenamefont {Lee},
  \citenamefont {Huo},\ and\ \citenamefont {Coker}}]{CokerAnnRevPhysChem16}%
  \BibitemOpen
  \bibfield  {author} {\bibinfo {author} {\bibfnamefont {M.~K.}\ \bibnamefont
  {Lee}}, \bibinfo {author} {\bibfnamefont {P.}~\bibnamefont {Huo}}, \ and\
  \bibinfo {author} {\bibfnamefont {D.~F.}\ \bibnamefont {Coker}},\ }\href
  {\doibase 10.1146/annurev-physchem-040215-112252} {\bibfield  {journal}
  {\bibinfo  {journal} {Ann. Rev. Phys. Chem.}\ }\textbf {\bibinfo {volume}
  {67}},\ \bibinfo {pages} {639} (\bibinfo {year} {2016})}\BibitemShut
  {NoStop}%
\bibitem [{\citenamefont {Kr{\" u}ger}\ \emph {et~al.}(2015)\citenamefont
  {Kr{\" u}ger}, \citenamefont {Bartels}, \citenamefont {Bartels},
  \citenamefont {Kandratsenka}, \citenamefont {Tully}, \citenamefont {Wodtke},\
  and\ \citenamefont {Sch{\" a}fer}}]{TullyWodtkeShaferJPCC15}%
  \BibitemOpen
  \bibfield  {author} {\bibinfo {author} {\bibfnamefont {B.~C.}\ \bibnamefont
  {Kr{\" u}ger}}, \bibinfo {author} {\bibfnamefont {N.}~\bibnamefont
  {Bartels}}, \bibinfo {author} {\bibfnamefont {C.}~\bibnamefont {Bartels}},
  \bibinfo {author} {\bibfnamefont {A.}~\bibnamefont {Kandratsenka}}, \bibinfo
  {author} {\bibfnamefont {J.~C.}\ \bibnamefont {Tully}}, \bibinfo {author}
  {\bibfnamefont {A.~M.}\ \bibnamefont {Wodtke}}, \ and\ \bibinfo {author}
  {\bibfnamefont {T.}~\bibnamefont {Sch{\" a}fer}},\ }\href {\doibase
  10.1021/acs.jpcc.5b00388} {\bibfield  {journal} {\bibinfo  {journal} {J.
  Phys. Chem. C}\ }\textbf {\bibinfo {volume} {119}},\ \bibinfo {pages} {3268}
  (\bibinfo {year} {2015})}\BibitemShut {NoStop}%
\bibitem [{\citenamefont {Tiwari}\ and\ \citenamefont
  {Henriksen}(2016)}]{HenricksenJCP16}%
  \BibitemOpen
  \bibfield  {author} {\bibinfo {author} {\bibfnamefont {A.~K.}\ \bibnamefont
  {Tiwari}}\ and\ \bibinfo {author} {\bibfnamefont {N.~E.}\ \bibnamefont
  {Henriksen}},\ }\href@noop {} {\bibfield  {journal} {\bibinfo  {journal} {J.
  Chem. Phys.}\ }\textbf {\bibinfo {volume} {144}},\ \bibinfo {eid} {014306}
  (\bibinfo {year} {2016})}\BibitemShut {NoStop}%
\bibitem [{\citenamefont {Miwa}\ \emph {et~al.}(2019)\citenamefont {Miwa},
  \citenamefont {Imada}, \citenamefont {Imai-Imada}, \citenamefont {Kimura},
  \citenamefont {Galperin},\ and\ \citenamefont {Kim}}]{ourNL}%
  \BibitemOpen
  \bibfield  {author} {\bibinfo {author} {\bibfnamefont {K.}~\bibnamefont
  {Miwa}}, \bibinfo {author} {\bibfnamefont {H.}~\bibnamefont {Imada}},
  \bibinfo {author} {\bibfnamefont {M.}~\bibnamefont {Imai-Imada}}, \bibinfo
  {author} {\bibfnamefont {K.}~\bibnamefont {Kimura}}, \bibinfo {author}
  {\bibfnamefont {M.}~\bibnamefont {Galperin}}, \ and\ \bibinfo {author}
  {\bibfnamefont {Y.}~\bibnamefont {Kim}},\ }\href {\doibase
  10.1021/acs.nanolett.8b04484} {\bibfield  {journal} {\bibinfo  {journal}
  {Nano Lett.}\ } (\bibinfo {year} {2019}),\ 10.1021/acs.nanolett.8b04484},\
  \Eprint {http://arxiv.org/abs/https://doi.org/10.1021/acs.nanolett.8b04484}
  {https://doi.org/10.1021/acs.nanolett.8b04484} \BibitemShut {NoStop}%
\bibitem [{\citenamefont {Long}\ \emph {et~al.}(2016)\citenamefont {Long},
  \citenamefont {Guo}, \citenamefont {Liu},\ and\ \citenamefont
  {Fang}}]{LongJPCL16}%
  \BibitemOpen
  \bibfield  {author} {\bibinfo {author} {\bibfnamefont {R.}~\bibnamefont
  {Long}}, \bibinfo {author} {\bibfnamefont {M.}~\bibnamefont {Guo}}, \bibinfo
  {author} {\bibfnamefont {L.}~\bibnamefont {Liu}}, \ and\ \bibinfo {author}
  {\bibfnamefont {W.}~\bibnamefont {Fang}},\ }\href {\doibase
  10.1021/acs.jpclett.6b00757} {\bibfield  {journal} {\bibinfo  {journal} {J.
  Phys. Chem. Lett.}\ }\textbf {\bibinfo {volume} {7}},\ \bibinfo {pages}
  {1830} (\bibinfo {year} {2016})}\BibitemShut {NoStop}%
\bibitem [{\citenamefont {Nitzan}\ and\ \citenamefont
  {Galperin}(2018)}]{nitzan_kinetic_2018}%
  \BibitemOpen
  \bibfield  {author} {\bibinfo {author} {\bibfnamefont {A.}~\bibnamefont
  {Nitzan}}\ and\ \bibinfo {author} {\bibfnamefont {M.}~\bibnamefont
  {Galperin}},\ }\href {\doibase 10.1021/acs.jpclett.8b01886} {\bibfield
  {journal} {\bibinfo  {journal} {J. Phys. Chem. Lett.}\ }\textbf {\bibinfo
  {volume} {9}},\ \bibinfo {pages} {4886} (\bibinfo {year} {2018})}\BibitemShut
  {NoStop}%
\bibitem [{\citenamefont {Gelbwaser-Klimovsky}\ \emph
  {et~al.}(2018)\citenamefont {Gelbwaser-Klimovsky}, \citenamefont
  {Aspuru-Guzik}, \citenamefont {Thoss},\ and\ \citenamefont
  {Peskin}}]{ThossPeskinNL18}%
  \BibitemOpen
  \bibfield  {author} {\bibinfo {author} {\bibfnamefont {D.}~\bibnamefont
  {Gelbwaser-Klimovsky}}, \bibinfo {author} {\bibfnamefont {A.}~\bibnamefont
  {Aspuru-Guzik}}, \bibinfo {author} {\bibfnamefont {M.}~\bibnamefont {Thoss}},
  \ and\ \bibinfo {author} {\bibfnamefont {U.}~\bibnamefont {Peskin}},\ }\href
  {\doibase 10.1021/acs.nanolett.8b01127} {\bibfield  {journal} {\bibinfo
  {journal} {Nano Lett.}\ }\textbf {\bibinfo {volume} {18}},\ \bibinfo {pages}
  {4727} (\bibinfo {year} {2018})},\ \Eprint
  {http://arxiv.org/abs/https://doi.org/10.1021/acs.nanolett.8b01127}
  {https://doi.org/10.1021/acs.nanolett.8b01127} \BibitemShut {NoStop}%
\bibitem [{\citenamefont {Foti}\ and\ \citenamefont {V{\'
  a}zquez}(2018)}]{FortiVazquezJPCL18}%
  \BibitemOpen
  \bibfield  {author} {\bibinfo {author} {\bibfnamefont {G.}~\bibnamefont
  {Foti}}\ and\ \bibinfo {author} {\bibfnamefont {H.}~\bibnamefont {V{\'
  a}zquez}},\ }\href {\doibase 10.1021/acs.jpclett.8b00940} {\bibfield
  {journal} {\bibinfo  {journal} {J. Phys. Chem. Lett.}\ }\textbf {\bibinfo
  {volume} {9}},\ \bibinfo {pages} {2791} (\bibinfo {year} {2018})}\BibitemShut
  {NoStop}%
\bibitem [{\citenamefont {Simine}\ and\ \citenamefont
  {Segal}(2012)}]{SegalPCCP12}%
  \BibitemOpen
  \bibfield  {author} {\bibinfo {author} {\bibfnamefont {L.}~\bibnamefont
  {Simine}}\ and\ \bibinfo {author} {\bibfnamefont {D.}~\bibnamefont {Segal}},\
  }\href {\doibase 10.1039/C2CP40851A} {\bibfield  {journal} {\bibinfo
  {journal} {Phys. Chem. Chem. Phys.}\ }\textbf {\bibinfo {volume} {14}},\
  \bibinfo {pages} {13820} (\bibinfo {year} {2012})}\BibitemShut {NoStop}%
\bibitem [{\citenamefont {L\"u}\ \emph {et~al.}(2011)\citenamefont {L\"u},
  \citenamefont {Hedeg\aa{}rd},\ and\ \citenamefont
  {Brandbyge}}]{BrandbygePRL11}%
  \BibitemOpen
  \bibfield  {author} {\bibinfo {author} {\bibfnamefont {J.-T.}\ \bibnamefont
  {L\"u}}, \bibinfo {author} {\bibfnamefont {P.}~\bibnamefont {Hedeg\aa{}rd}},
  \ and\ \bibinfo {author} {\bibfnamefont {M.}~\bibnamefont {Brandbyge}},\
  }\href {\doibase 10.1103/PhysRevLett.107.046801} {\bibfield  {journal}
  {\bibinfo  {journal} {Phys. Rev. Lett.}\ }\textbf {\bibinfo {volume} {107}},\
  \bibinfo {pages} {046801} (\bibinfo {year} {2011})}\BibitemShut {NoStop}%
\bibitem [{\citenamefont {Ohmura}\ \emph {et~al.}(2016)\citenamefont {Ohmura},
  \citenamefont {Tsuruta}, \citenamefont {Shimojo},\ and\ \citenamefont
  {Nakano}}]{OhmuraAIPAdv16}%
  \BibitemOpen
  \bibfield  {author} {\bibinfo {author} {\bibfnamefont {S.}~\bibnamefont
  {Ohmura}}, \bibinfo {author} {\bibfnamefont {K.}~\bibnamefont {Tsuruta}},
  \bibinfo {author} {\bibfnamefont {F.}~\bibnamefont {Shimojo}}, \ and\
  \bibinfo {author} {\bibfnamefont {A.}~\bibnamefont {Nakano}},\ }\href@noop {}
  {\bibfield  {journal} {\bibinfo  {journal} {AIP Advances}\ }\textbf {\bibinfo
  {volume} {6}},\ \bibinfo {eid} {015305} (\bibinfo {year} {2016})}\BibitemShut
  {NoStop}%
\bibitem [{\citenamefont {Nelson}\ \emph {et~al.}(2014)\citenamefont {Nelson},
  \citenamefont {Fernandez-Alberti}, \citenamefont {Roitberg},\ and\
  \citenamefont {Tretiak}}]{TretiakAccChemRes14}%
  \BibitemOpen
  \bibfield  {author} {\bibinfo {author} {\bibfnamefont {T.}~\bibnamefont
  {Nelson}}, \bibinfo {author} {\bibfnamefont {S.}~\bibnamefont
  {Fernandez-Alberti}}, \bibinfo {author} {\bibfnamefont {A.~E.}\ \bibnamefont
  {Roitberg}}, \ and\ \bibinfo {author} {\bibfnamefont {S.}~\bibnamefont
  {Tretiak}},\ }\href {\doibase 10.1021/ar400263p} {\bibfield  {journal}
  {\bibinfo  {journal} {Acc. Chem. Res.}\ }\textbf {\bibinfo {volume} {47}},\
  \bibinfo {pages} {1155} (\bibinfo {year} {2014})}\BibitemShut {NoStop}%
\bibitem [{\citenamefont {Nikiforov}\ \emph {et~al.}(2016)\citenamefont
  {Nikiforov}, \citenamefont {Gamez}, \citenamefont {Thiel},\ and\
  \citenamefont {Filatov}}]{FilatovJPCL16}%
  \BibitemOpen
  \bibfield  {author} {\bibinfo {author} {\bibfnamefont {A.}~\bibnamefont
  {Nikiforov}}, \bibinfo {author} {\bibfnamefont {J.~A.}\ \bibnamefont
  {Gamez}}, \bibinfo {author} {\bibfnamefont {W.}~\bibnamefont {Thiel}}, \ and\
  \bibinfo {author} {\bibfnamefont {M.}~\bibnamefont {Filatov}},\ }\href
  {\doibase 10.1021/acs.jpclett.5b02575} {\bibfield  {journal} {\bibinfo
  {journal} {J. Phys. Chem. Lett.}\ }\textbf {\bibinfo {volume} {7}},\ \bibinfo
  {pages} {105} (\bibinfo {year} {2016})}\BibitemShut {NoStop}%
\bibitem [{\citenamefont {Brandbyge}\ \emph {et~al.}(1995)\citenamefont
  {Brandbyge}, \citenamefont {Hedeg\aa{}rd}, \citenamefont {Heinz},
  \citenamefont {Misewich},\ and\ \citenamefont {Newns}}]{NewnsPRB95}%
  \BibitemOpen
  \bibfield  {author} {\bibinfo {author} {\bibfnamefont {M.}~\bibnamefont
  {Brandbyge}}, \bibinfo {author} {\bibfnamefont {P.}~\bibnamefont
  {Hedeg\aa{}rd}}, \bibinfo {author} {\bibfnamefont {T.~F.}\ \bibnamefont
  {Heinz}}, \bibinfo {author} {\bibfnamefont {J.~A.}\ \bibnamefont {Misewich}},
  \ and\ \bibinfo {author} {\bibfnamefont {D.~M.}\ \bibnamefont {Newns}},\
  }\href {\doibase 10.1103/PhysRevB.52.6042} {\bibfield  {journal} {\bibinfo
  {journal} {Phys. Rev. B}\ }\textbf {\bibinfo {volume} {52}},\ \bibinfo
  {pages} {6042} (\bibinfo {year} {1995})}\BibitemShut {NoStop}%
\bibitem [{\citenamefont {L\"u}\ \emph {et~al.}(2012)\citenamefont {L\"u},
  \citenamefont {Brandbyge}, \citenamefont {Hedeg\aa{}rd}, \citenamefont
  {Todorov},\ and\ \citenamefont {Dundas}}]{TodorovPRB12}%
  \BibitemOpen
  \bibfield  {author} {\bibinfo {author} {\bibfnamefont {J.-T.}\ \bibnamefont
  {L\"u}}, \bibinfo {author} {\bibfnamefont {M.}~\bibnamefont {Brandbyge}},
  \bibinfo {author} {\bibfnamefont {P.}~\bibnamefont {Hedeg\aa{}rd}}, \bibinfo
  {author} {\bibfnamefont {T.~N.}\ \bibnamefont {Todorov}}, \ and\ \bibinfo
  {author} {\bibfnamefont {D.}~\bibnamefont {Dundas}},\ }\href {\doibase
  10.1103/PhysRevB.85.245444} {\bibfield  {journal} {\bibinfo  {journal} {Phys.
  Rev. B}\ }\textbf {\bibinfo {volume} {85}},\ \bibinfo {pages} {245444}
  (\bibinfo {year} {2012})}\BibitemShut {NoStop}%
\bibitem [{\citenamefont {L\"u}\ \emph {et~al.}(2015)\citenamefont {L\"u},
  \citenamefont {Christensen}, \citenamefont {Wang}, \citenamefont
  {Hedeg\aa{}rd},\ and\ \citenamefont {Brandbyge}}]{HedegardBrandbygePRL15}%
  \BibitemOpen
  \bibfield  {author} {\bibinfo {author} {\bibfnamefont {J.-T.}\ \bibnamefont
  {L\"u}}, \bibinfo {author} {\bibfnamefont {R.~B.}\ \bibnamefont
  {Christensen}}, \bibinfo {author} {\bibfnamefont {J.-S.}\ \bibnamefont
  {Wang}}, \bibinfo {author} {\bibfnamefont {P.}~\bibnamefont {Hedeg\aa{}rd}},
  \ and\ \bibinfo {author} {\bibfnamefont {M.}~\bibnamefont {Brandbyge}},\
  }\href {\doibase 10.1103/PhysRevLett.114.096801} {\bibfield  {journal}
  {\bibinfo  {journal} {Phys. Rev. Lett.}\ }\textbf {\bibinfo {volume} {114}},\
  \bibinfo {pages} {096801} (\bibinfo {year} {2015})}\BibitemShut {NoStop}%
\bibitem [{\citenamefont {Thomas}\ \emph {et~al.}(2012)\citenamefont {Thomas},
  \citenamefont {Karzig}, \citenamefont {Kusminskiy}, \citenamefont
  {Zar\'and},\ and\ \citenamefont {von Oppen}}]{VonOppenPRB12}%
  \BibitemOpen
  \bibfield  {author} {\bibinfo {author} {\bibfnamefont {M.}~\bibnamefont
  {Thomas}}, \bibinfo {author} {\bibfnamefont {T.}~\bibnamefont {Karzig}},
  \bibinfo {author} {\bibfnamefont {S.~V.}\ \bibnamefont {Kusminskiy}},
  \bibinfo {author} {\bibfnamefont {G.}~\bibnamefont {Zar\'and}}, \ and\
  \bibinfo {author} {\bibfnamefont {F.}~\bibnamefont {von Oppen}},\ }\href
  {\doibase 10.1103/PhysRevB.86.195419} {\bibfield  {journal} {\bibinfo
  {journal} {Phys. Rev. B}\ }\textbf {\bibinfo {volume} {86}},\ \bibinfo
  {pages} {195419} (\bibinfo {year} {2012})}\BibitemShut {NoStop}%
\bibitem [{\citenamefont {Bode}\ \emph {et~al.}(2011)\citenamefont {Bode},
  \citenamefont {Kusminskiy}, \citenamefont {Egger},\ and\ \citenamefont {von
  Oppen}}]{vonOppenPRL11}%
  \BibitemOpen
  \bibfield  {author} {\bibinfo {author} {\bibfnamefont {N.}~\bibnamefont
  {Bode}}, \bibinfo {author} {\bibfnamefont {S.~V.}\ \bibnamefont
  {Kusminskiy}}, \bibinfo {author} {\bibfnamefont {R.}~\bibnamefont {Egger}}, \
  and\ \bibinfo {author} {\bibfnamefont {F.}~\bibnamefont {von Oppen}},\ }\href
  {\doibase 10.1103/PhysRevLett.107.036804} {\bibfield  {journal} {\bibinfo
  {journal} {Phys. Rev. Lett.}\ }\textbf {\bibinfo {volume} {107}},\ \bibinfo
  {pages} {036804} (\bibinfo {year} {2011})}\BibitemShut {NoStop}%
\bibitem [{\citenamefont {Bode}\ \emph {et~al.}(2012)\citenamefont {Bode},
  \citenamefont {Kusminskiy}, \citenamefont {Egger},\ and\ \citenamefont {von
  Oppen}}]{vonOppenBJN12}%
  \BibitemOpen
  \bibfield  {author} {\bibinfo {author} {\bibfnamefont {N.}~\bibnamefont
  {Bode}}, \bibinfo {author} {\bibfnamefont {S.~V.}\ \bibnamefont
  {Kusminskiy}}, \bibinfo {author} {\bibfnamefont {R.~E.}\ \bibnamefont
  {Egger}}, \ and\ \bibinfo {author} {\bibfnamefont {F.}~\bibnamefont {von
  Oppen}},\ }\href {\doibase 10.3762/bjnano.3.15} {\bibfield  {journal}
  {\bibinfo  {journal} {Beilstein J. Nanotechnol.}\ }\textbf {\bibinfo {volume}
  {3}},\ \bibinfo {pages} {144–162} (\bibinfo {year} {2012})}\BibitemShut
  {NoStop}%
\bibitem [{\citenamefont {Shenvi}\ and\ \citenamefont
  {Tully}(2012)}]{ShenviTullyFaradDisc12}%
  \BibitemOpen
  \bibfield  {author} {\bibinfo {author} {\bibfnamefont {N.}~\bibnamefont
  {Shenvi}}\ and\ \bibinfo {author} {\bibfnamefont {J.~C.}\ \bibnamefont
  {Tully}},\ }\href {\doibase 10.1039/C2FD20032E} {\bibfield  {journal}
  {\bibinfo  {journal} {Faraday Discuss.}\ }\textbf {\bibinfo {volume} {157}},\
  \bibinfo {pages} {325} (\bibinfo {year} {2012})}\BibitemShut {NoStop}%
\bibitem [{\citenamefont {Falk}\ \emph {et~al.}(2014)\citenamefont {Falk},
  \citenamefont {Landry},\ and\ \citenamefont {Subotnik}}]{SubotnikJPCB14}%
  \BibitemOpen
  \bibfield  {author} {\bibinfo {author} {\bibfnamefont {M.~J.}\ \bibnamefont
  {Falk}}, \bibinfo {author} {\bibfnamefont {B.~R.}\ \bibnamefont {Landry}}, \
  and\ \bibinfo {author} {\bibfnamefont {J.~E.}\ \bibnamefont {Subotnik}},\
  }\href {\doibase 10.1021/jp5011346} {\bibfield  {journal} {\bibinfo
  {journal} {J. Phys. Chem. B}\ }\textbf {\bibinfo {volume} {118}},\ \bibinfo
  {pages} {8108} (\bibinfo {year} {2014})}\BibitemShut {NoStop}%
\bibitem [{\citenamefont {Dou}\ \emph {et~al.}(2015)\citenamefont {Dou},
  \citenamefont {Nitzan},\ and\ \citenamefont
  {Subotnik}}]{NitzanSubotnikJCP15}%
  \BibitemOpen
  \bibfield  {author} {\bibinfo {author} {\bibfnamefont {W.}~\bibnamefont
  {Dou}}, \bibinfo {author} {\bibfnamefont {A.}~\bibnamefont {Nitzan}}, \ and\
  \bibinfo {author} {\bibfnamefont {J.~E.}\ \bibnamefont {Subotnik}},\
  }\href@noop {} {\bibfield  {journal} {\bibinfo  {journal} {J. Chem. Phys.}\
  }\textbf {\bibinfo {volume} {143}},\ \bibinfo {eid} {054103} (\bibinfo {year}
  {2015})}\BibitemShut {NoStop}%
\bibitem [{\citenamefont {Dou}\ and\ \citenamefont
  {Subotnik}(2016)}]{SubotnikJCP16}%
  \BibitemOpen
  \bibfield  {author} {\bibinfo {author} {\bibfnamefont {W.}~\bibnamefont
  {Dou}}\ and\ \bibinfo {author} {\bibfnamefont {J.~E.}\ \bibnamefont
  {Subotnik}},\ }\href@noop {} {\bibfield  {journal} {\bibinfo  {journal} {J.
  Chem. Phys.}\ }\textbf {\bibinfo {volume} {145}},\ \bibinfo {eid} {054102}
  (\bibinfo {year} {2016})}\BibitemShut {NoStop}%
\bibitem [{\citenamefont {Dou}\ \emph {et~al.}(2017)\citenamefont {Dou},
  \citenamefont {Miao},\ and\ \citenamefont {Subotnik}}]{SubotnikPRL17}%
  \BibitemOpen
  \bibfield  {author} {\bibinfo {author} {\bibfnamefont {W.}~\bibnamefont
  {Dou}}, \bibinfo {author} {\bibfnamefont {G.}~\bibnamefont {Miao}}, \ and\
  \bibinfo {author} {\bibfnamefont {J.~E.}\ \bibnamefont {Subotnik}},\ }\href
  {\doibase 10.1103/PhysRevLett.119.046001} {\bibfield  {journal} {\bibinfo
  {journal} {Phys. Rev. Lett.}\ }\textbf {\bibinfo {volume} {119}},\ \bibinfo
  {pages} {046001} (\bibinfo {year} {2017})}\BibitemShut {NoStop}%
\bibitem [{\citenamefont {Dou}\ and\ \citenamefont
  {Subotnik}(2017)}]{SubotnikPRB17}%
  \BibitemOpen
  \bibfield  {author} {\bibinfo {author} {\bibfnamefont {W.}~\bibnamefont
  {Dou}}\ and\ \bibinfo {author} {\bibfnamefont {J.~E.}\ \bibnamefont
  {Subotnik}},\ }\href {\doibase 10.1103/PhysRevB.96.104305} {\bibfield
  {journal} {\bibinfo  {journal} {Phys. Rev. B}\ }\textbf {\bibinfo {volume}
  {96}},\ \bibinfo {pages} {104305} (\bibinfo {year} {2017})}\BibitemShut
  {NoStop}%
\bibitem [{\citenamefont {Askerka}\ \emph {et~al.}(2016)\citenamefont
  {Askerka}, \citenamefont {Maurer}, \citenamefont {Batista},\ and\
  \citenamefont {Tully}}]{TullyPRL16}%
  \BibitemOpen
  \bibfield  {author} {\bibinfo {author} {\bibfnamefont {M.}~\bibnamefont
  {Askerka}}, \bibinfo {author} {\bibfnamefont {R.~J.}\ \bibnamefont {Maurer}},
  \bibinfo {author} {\bibfnamefont {V.~S.}\ \bibnamefont {Batista}}, \ and\
  \bibinfo {author} {\bibfnamefont {J.~C.}\ \bibnamefont {Tully}},\ }\href
  {\doibase 10.1103/PhysRevLett.116.217601} {\bibfield  {journal} {\bibinfo
  {journal} {Phys. Rev. Lett.}\ }\textbf {\bibinfo {volume} {116}},\ \bibinfo
  {pages} {217601} (\bibinfo {year} {2016})}\BibitemShut {NoStop}%
\bibitem [{\citenamefont {Maurer}\ \emph {et~al.}(2016)\citenamefont {Maurer},
  \citenamefont {Askerka}, \citenamefont {Batista},\ and\ \citenamefont
  {Tully}}]{TullyPRB16}%
  \BibitemOpen
  \bibfield  {author} {\bibinfo {author} {\bibfnamefont {R.~J.}\ \bibnamefont
  {Maurer}}, \bibinfo {author} {\bibfnamefont {M.}~\bibnamefont {Askerka}},
  \bibinfo {author} {\bibfnamefont {V.~S.}\ \bibnamefont {Batista}}, \ and\
  \bibinfo {author} {\bibfnamefont {J.~C.}\ \bibnamefont {Tully}},\ }\href
  {\doibase 10.1103/PhysRevB.94.115432} {\bibfield  {journal} {\bibinfo
  {journal} {Phys. Rev. B}\ }\textbf {\bibinfo {volume} {94}},\ \bibinfo
  {pages} {115432} (\bibinfo {year} {2016})}\BibitemShut {NoStop}%
\bibitem [{\citenamefont {Hopjan}\ \emph {et~al.}(2018)\citenamefont {Hopjan},
  \citenamefont {Stefanucci}, \citenamefont {Perfetto},\ and\ \citenamefont
  {Verdozzi}}]{hopjan_molecular_2018}%
  \BibitemOpen
  \bibfield  {author} {\bibinfo {author} {\bibfnamefont {M.}~\bibnamefont
  {Hopjan}}, \bibinfo {author} {\bibfnamefont {G.}~\bibnamefont {Stefanucci}},
  \bibinfo {author} {\bibfnamefont {E.}~\bibnamefont {Perfetto}}, \ and\
  \bibinfo {author} {\bibfnamefont {C.}~\bibnamefont {Verdozzi}},\ }\href
  {https://link.aps.org/doi/10.1103/PhysRevB.98.041405} {\bibfield  {journal}
  {\bibinfo  {journal} {Phys. Rev. B}\ }\textbf {\bibinfo {volume} {98}},\
  \bibinfo {pages} {041405(R)} (\bibinfo {year} {2018})}\BibitemShut {NoStop}%
\bibitem [{\citenamefont {Kantorovich}(2018)}]{kantorovich_nonadiabatic_2018}%
  \BibitemOpen
  \bibfield  {author} {\bibinfo {author} {\bibfnamefont {L.}~\bibnamefont
  {Kantorovich}},\ }\href {https://link.aps.org/doi/10.1103/PhysRevB.98.014307}
  {\bibfield  {journal} {\bibinfo  {journal} {Phys. Rev. B}\ }\textbf {\bibinfo
  {volume} {98}},\ \bibinfo {pages} {014307} (\bibinfo {year}
  {2018})}\BibitemShut {NoStop}%
\bibitem [{\citenamefont {Chen}\ \emph {et~al.}(2019)\citenamefont {Chen},
  \citenamefont {Miwa},\ and\ \citenamefont
  {Galperin}}]{chen_current-induced_2019}%
  \BibitemOpen
  \bibfield  {author} {\bibinfo {author} {\bibfnamefont {F.}~\bibnamefont
  {Chen}}, \bibinfo {author} {\bibfnamefont {K.}~\bibnamefont {Miwa}}, \ and\
  \bibinfo {author} {\bibfnamefont {M.}~\bibnamefont {Galperin}},\ }\href
  {\doibase 10.1021/acs.jpca.8b09251} {\bibfield  {journal} {\bibinfo
  {journal} {J. Phys. Chem. A}\ }\textbf {\bibinfo {volume} {123}},\ \bibinfo
  {pages} {693} (\bibinfo {year} {2019})}\BibitemShut {NoStop}%
\bibitem [{\citenamefont {Chen}\ \emph {et~al.}(2017)\citenamefont {Chen},
  \citenamefont {Ochoa},\ and\ \citenamefont {Galperin}}]{ChenOchoaMGJCP17}%
  \BibitemOpen
  \bibfield  {author} {\bibinfo {author} {\bibfnamefont {F.}~\bibnamefont
  {Chen}}, \bibinfo {author} {\bibfnamefont {M.~A.}\ \bibnamefont {Ochoa}}, \
  and\ \bibinfo {author} {\bibfnamefont {M.}~\bibnamefont {Galperin}},\ }\href
  {\doibase http://dx.doi.org/10.1063/1.4965825} {\bibfield  {journal}
  {\bibinfo  {journal} {J. Chem. Phys.}\ }\textbf {\bibinfo {volume} {146}},\
  \bibinfo {pages} {092301} (\bibinfo {year} {2017})}\BibitemShut {NoStop}%
\bibitem [{\citenamefont {Miwa}\ \emph {et~al.}(2017)\citenamefont {Miwa},
  \citenamefont {Chen},\ and\ \citenamefont {Galperin}}]{MiwaChenMGSciRep17}%
  \BibitemOpen
  \bibfield  {author} {\bibinfo {author} {\bibfnamefont {K.}~\bibnamefont
  {Miwa}}, \bibinfo {author} {\bibfnamefont {F.}~\bibnamefont {Chen}}, \ and\
  \bibinfo {author} {\bibfnamefont {M.}~\bibnamefont {Galperin}},\ }\href
  {\doibase 10.1038/s41598-017-09060-0} {\bibfield  {journal} {\bibinfo
  {journal} {Sci. Rep.}\ }\textbf {\bibinfo {volume} {7}},\ \bibinfo {pages}
  {9735} (\bibinfo {year} {2017})}\BibitemShut {NoStop}%
\bibitem [{\citenamefont {Kongsuwan}\ \emph {et~al.}(2018)\citenamefont
  {Kongsuwan}, \citenamefont {Demetriadou}, \citenamefont {Chikkaraddy},
  \citenamefont {Benz}, \citenamefont {Turek}, \citenamefont {Keyser},
  \citenamefont {Baumberg},\ and\ \citenamefont
  {Hess}}]{kongsuwan_suppressed_2018}%
  \BibitemOpen
  \bibfield  {author} {\bibinfo {author} {\bibfnamefont {N.}~\bibnamefont
  {Kongsuwan}}, \bibinfo {author} {\bibfnamefont {A.}~\bibnamefont
  {Demetriadou}}, \bibinfo {author} {\bibfnamefont {R.}~\bibnamefont
  {Chikkaraddy}}, \bibinfo {author} {\bibfnamefont {F.}~\bibnamefont {Benz}},
  \bibinfo {author} {\bibfnamefont {V.~A.}\ \bibnamefont {Turek}}, \bibinfo
  {author} {\bibfnamefont {U.~F.}\ \bibnamefont {Keyser}}, \bibinfo {author}
  {\bibfnamefont {J.~J.}\ \bibnamefont {Baumberg}}, \ and\ \bibinfo {author}
  {\bibfnamefont {O.}~\bibnamefont {Hess}},\ }\href {\doibase
  10.1021/acsphotonics.7b00668} {\bibfield  {journal} {\bibinfo  {journal} {ACS
  Photonics}\ }\textbf {\bibinfo {volume} {5}},\ \bibinfo {pages} {186}
  (\bibinfo {year} {2018})}\BibitemShut {NoStop}%
\bibitem [{\citenamefont {Chikkaraddy}\ \emph {et~al.}(2016)\citenamefont
  {Chikkaraddy}, \citenamefont {de~Nijs}, \citenamefont {Benz}, \citenamefont
  {Barrow}, \citenamefont {Scherman}, \citenamefont {Rosta}, \citenamefont
  {Demetriadou}, \citenamefont {Fox}, \citenamefont {Hess},\ and\ \citenamefont
  {Baumberg}}]{chikkaraddy_single-molecule_2016}%
  \BibitemOpen
  \bibfield  {author} {\bibinfo {author} {\bibfnamefont {R.}~\bibnamefont
  {Chikkaraddy}}, \bibinfo {author} {\bibfnamefont {B.}~\bibnamefont
  {de~Nijs}}, \bibinfo {author} {\bibfnamefont {F.}~\bibnamefont {Benz}},
  \bibinfo {author} {\bibfnamefont {S.~J.}\ \bibnamefont {Barrow}}, \bibinfo
  {author} {\bibfnamefont {O.~A.}\ \bibnamefont {Scherman}}, \bibinfo {author}
  {\bibfnamefont {E.}~\bibnamefont {Rosta}}, \bibinfo {author} {\bibfnamefont
  {A.}~\bibnamefont {Demetriadou}}, \bibinfo {author} {\bibfnamefont
  {P.}~\bibnamefont {Fox}}, \bibinfo {author} {\bibfnamefont {O.}~\bibnamefont
  {Hess}}, \ and\ \bibinfo {author} {\bibfnamefont {J.~J.}\ \bibnamefont
  {Baumberg}},\ }\href {\doibase 10.1038/nature17974} {\bibfield  {journal}
  {\bibinfo  {journal} {Nature}\ }\textbf {\bibinfo {volume} {535}},\ \bibinfo
  {pages} {127} (\bibinfo {year} {2016})}\BibitemShut {NoStop}%
\bibitem [{Note1()}]{Note1}%
  \BibitemOpen
  \bibinfo {note} {Note that original Head-Gordon and Tully expression for
  electronic friction~\cite {TullyJCP95} comes from equilibrium quasiparticle
  consideration of the mentioned subset of terms.}\BibitemShut {Stop}%
\bibitem [{\citenamefont {Head-Gordon}\ and\ \citenamefont
  {Tully}(1995)}]{TullyJCP95}%
  \BibitemOpen
  \bibfield  {author} {\bibinfo {author} {\bibfnamefont {M.}~\bibnamefont
  {Head-Gordon}}\ and\ \bibinfo {author} {\bibfnamefont {J.~C.}\ \bibnamefont
  {Tully}},\ }\href {\doibase 10.1063/1.469915} {\bibfield  {journal} {\bibinfo
   {journal} {J. Chem. Phys.}\ }\textbf {\bibinfo {volume} {103}},\ \bibinfo
  {pages} {10137} (\bibinfo {year} {1995})}\BibitemShut {NoStop}%
\end{thebibliography}

%

\end{document}